\documentclass[prl,aps,twocolumn,showpacs,amsmath,amssymb,floatfix,citeautoscript]{revtex4-1}
\usepackage{graphicx}
\usepackage{dcolumn}
\usepackage{bm}
\usepackage{amsmath}
\usepackage{amssymb}
\usepackage{amssymb, verbatim}
\usepackage[normal]{subfigure}

\usepackage{hyperref}
\hypersetup{
        colorlinks=true,
}

\makeatletter

\usepackage{color}

\begin{document}

\title{Probing Majorana Physics in Quantum Dot Shot Noise Experiments}

\author{Dong E. Liu, Meng Cheng, and Roman M. Lutchyn}
\affiliation{Microsoft Research, Station Q, University of California, Santa Barbara, CA, 93106, USA}

\date{\today}

\begin{abstract}
We consider a quantum dot coupled to a topological superconductor and two normal leads and study transport properties of the system. Using Keldysh path-integral approach, we study current fluctuations (shot noise) within the low-energy effective theory. We argue that the combination of the tunneling conductance and the shot noise through
a quantum dot allows one to distinguish between the topological (Majorana) and non-topological (e.g., Kondo) origin of the zero-bias conduction peak. Specifically, we show that, while the tunneling conductance might exhibit zero-bias anomaly due to Majorana or Kondo physics, the shot noise is qualitatively different in the presence of Majorana zero modes.
\end{abstract}

\pacs{73.21.Hb, 71.10.Pm, 74.78.Fk, 72.70.+m}

\maketitle

{\em Introduction}. The search for topological superconductors hosting non-Abelian quasiparticles (defects binding Majorana zero modes) has become an active and exciting pursuit in condensed matter physics\cite{Reich,Brouwer_Science,Wilczek2012}. There has been enormous theoretical and experimental activity in this direction recently~\cite{AliceaRev} fueled, in part, by the potential application of topological superconductors for the fault-tolerant topological quantum computation schemes\cite{TQCreview}.
%due to their exotic non-Abelian braiding statistical properties \cite{Moore1991,Nayak1996,ReadGreen} and potential application in topological quantum computation \cite{TQCreview}.
A large number of theoretical proposals for engineering topological superconductors in the laboratory has been put
forward  \cite{Fu&Kane08,Fu&Kane09,Sau10,Alicea10,LutchynPRL10,1DwiresOreg,Linder10,MajoranaTInanowires,SauNature'12,Nadj-Perge13}, and there has been a significant amount of experimental activity in this area recently~\cite{Mourik2012, Rokhinson2012, Das2012, Deng2012, Fink2012, Churchill2013, Chang_PRL2013, Kondo_Aguado,Lee_arxiv2013, Deng_arxiv2014}. One of the simplest ways to detect the presence of Majorana zero modes (MZMs) in topological superconductors (TSC) is tunneling spectroscopy. Indeed, the presence of MZMs leads to a quantized zero-bias conductance $G=2e^2/h$ \cite{ZeroBiasAnomaly0, Law09,ZeroBiasAnomaly31, FlensbergPRB10, 1DwiresLutchyn2,  ZeroBiasAnomaly61, Fidkowski2012, lutchyn_andreev13}.
The pioneering Majorana experiment based on a semiconductor/superconductor heterostructure
proposal \cite{LutchynPRL10,1DwiresOreg} was performed in Delft \cite{Mourik2012} 
where the observation of zero-bias peak in a finite magnetic field was reported, 
consistent with the theoretical predictions\cite{ZeroBiasAnomaly61}. However, other effects might also lead to the zero-bias anomaly
which spurred the debate\cite{Bagrets12,liujie12,Pikulin12,Rainis13,Neven13,LobosPRL13,Sau&DasSarma13,Hui14} 
as to the precise origin of the (un-quantized)
zero-bias conduction peak observed in recent tunneling
experiments~\cite{Mourik2012, Rokhinson2012, Das2012, Deng2012, Fink2012, Churchill2013}.
Therefore, additional experiments testing other properties on
MZMs\cite{Rokhinson2012, Leijnse11,Liu11,Golub'11,LeePRB13,Vernek14,ChengPRX2014,
DasSarmainterferometry,Grosfeldinterferometry,Fu&Kandinterferometry,NilssonNoise,AkhmerovPRL09,Pikulin12,Rainis13,DissipativeMF,Valentini14} are necessary in order to reach a consensus.

\begin{figure}
\centering
\includegraphics[width=3.4in,clip]{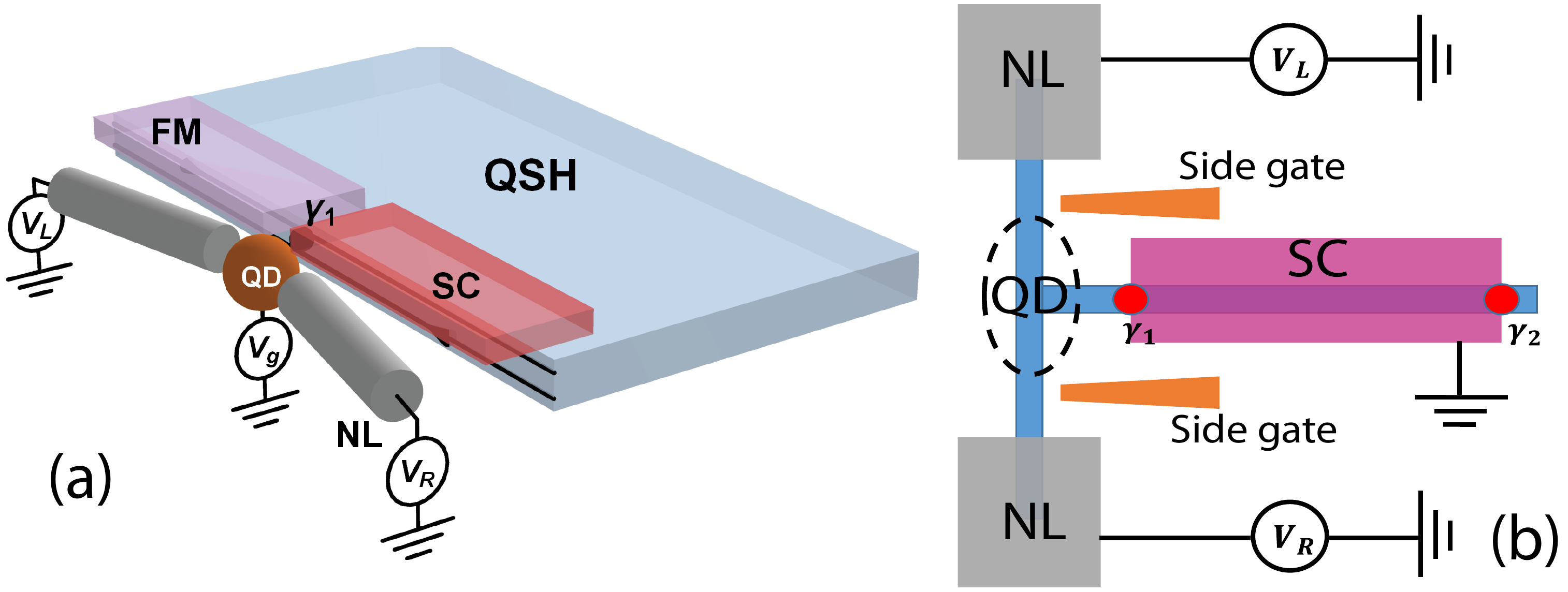}
\vspace{-0.1in}
\caption{Proposed experimental setup for the shot noise measurement. (a) A MZM
is formed at the domain wall between a ferromagnetic insulator and a s-wave superconductor at the QSH edge.
(b) A MZM is formed at the ends of topological superconducting wire, and QD is formed near the wire T-junction.
}
\vspace{-0.2in}
\label{fig:setup}
\end{figure}

In this Letter, we propose a new scheme, which combines the tunneling conductance and current fluctuation measurements,
to distinguish between the topological (Majorana) and non-topological origin of the zero-bias peak. 
We consider a quantum dot (QD) coupled to a MZM and two normal leads, see Fig. \ref{fig:setup}.
%In addition to the transport between a normal lead and a superconductor, 
One can extract information about MZMs by measuring the shot noise between two normal leads. 
This approach allows one to eliminate a number of false-positive features by simply changing either Majorana coupling or the QD couplings. 
Indeed, while the Kondo effect as well as resonant-tunneling physics exhibit zero-bias peaks in the tunneling conductance, their current fluctuations are {\it qualitatively} different. Thus, it is suggestive to use shot noise as a diagnostic tool for MZMs. The physics of the QD coupled to the MZM has been discussed in Refs.~\cite{Leijnse11, Golub'11, Lopez'2013, ChengPRX2014}. It has been shown that Majorana coupling significantly modifies the low-energy properties of the QD and drives the system to 
a new (different from Kondo) fixed point~\cite{ChengPRX2014}. 
Building on top of the slave boson formalism developed in Ref.~\cite{ChengPRX2014},
we compute here the shot noise in the system shown in Fig. \ref{fig:setup}. It is well-known that noise measurements usually provide
additional information for correlated systems \cite{shot_Fisher,TanakaPRB00, Meir02,SelaPRL06,BlanterRev} and often allow one to identify the nature of the charge carriers. The shot noise for the non-interacting systems such as the normal lead-TSC  and non-interacting QD-TSC have been considered in Refs.\cite{NilssonNoise,MajoranaShotNoise,MajoranaTunneling,ZocherPRL13,CaoSN2011}.
%However, we show that {\bf the effect of Coulomb interactions is very important due to the competition between Kondo and Majorana physics - add specific differences here}.
In this paper, we address this important and non-trivial question and obtain analytically the power spectrum of shot noise in the presence of the Coulomb interactions in QD by taking into account the interplay between Kondo and Majorana physics.

%\begin{widetext}
%\begin {table}[h]
%\begin{center}
%\caption{Table I: Shot noise power spectrum $P(\omega\rightarrow 0)$ and conductance $G$ \cite{Liu11,LeePRB13,Vernek14} for $\Gamma_L=\Gamma_R$.
%         }\label{tableI}
%    \begin{tabular}{ | p{4.0cm} | p{6.7cm} | p{6.8cm} | }
%    \hline
%       & with MZM  & without MZM \\ \hline
%    spinless system (non-interacting QD $U=0$) & $P(0)=\frac{e^2}{h}$ and $G=\frac{e^2}{2h}$,
%        (Do not depend on $\epsilon_d$) & $P(0)=0$ and $G=\frac{e^2}{h}$ for $\epsilon_d=0$, (Depend on $\epsilon_d$)\\ \hline
%    spinful system: Kondo regime ($|\epsilon_d|\gg \lambda,\Gamma$) & $P(0)=\frac{e^2}{h}$ and $G=\frac{3e^2}{2h}$
%                 & $P(0)=0$ and $G=\frac{2e^2}{h}$\\
%    \hline
%    \end{tabular}
%\end{center}
%\vspace{-0.2in}
%\end{table}
%\end{widetext}

Our main results are summarized in Table~\ref{tableI}. We find that in the case of symmetric couplings to the
leads $\Gamma_L=\Gamma_R$  the shot noise power spectrum $P(\omega\rightarrow 0)$ in the presence of MZM coupling
exhibits a universal value $e^2/2h$ which is independent of the QD energy level $\epsilon_d$,
and corresponds to the transmission probability $T(0)=\frac{1}{2}$ . This is to be contrasted with the resonant level model which exhibits a strong dependence on $\epsilon_d$.
In the Kondo limit, tunneling conductance exhibits the zero-bias anomaly but the shot noise power spectrum
is zero $P(\omega\rightarrow 0)=0$.
Thus, the combination of both shot noise and conductance through a QD allows one to distinguish between the
Majorana and other physics. We believe that our results are relevant for the ongoing Majorana experiments since
conductance and current fluctuations can be readily accessed.

\begin {table}
\begin{center}
\caption{Shot noise power spectrum $P(\omega\rightarrow 0)$ and conductance $G$ \cite{Liu11,LeePRB13,Vernek14} for $\Gamma_L=\Gamma_R$.
         }\label{tableI}
    \begin{tabular}{ | p{1.2cm} | p{3.6cm} | p{3.4cm} | }
    \hline
       & spinless system (non-interacting $U=0$)  & spinful system: Kondo regime ($|\epsilon_d|\gg \lambda,\Gamma$)\\ \hline
     with MZM & $P(0)=\frac{e^2}{2h}$ and $G=\frac{e^2}{2h}$,\newline
        (independent of $\epsilon_d$) & $P(0)=\frac{e^2}{2h}$ and $G=\frac{3e^2}{2h}$  \\ \hline
   without MZM &$P(0)=0$ and $G=\frac{e^2}{h}$ \newline for $\epsilon_d=0$
                 & $P(0)=0$ and $G=\frac{2e^2}{h}$\\
    \hline
    \end{tabular}
\end{center}
\vspace{-0.2in}
\end{table}

{\em Theoretical Model}. We consider a setup shown in Fig. \ref{fig:setup} in which a QD is coupled to a MZM $\gamma_1$ localized
at the domain wall between a magnetic insulator and an
s-wave superconductor at the edge of a Quantum Spin Hall (QSH) insulator \cite{Fu&Kane09} or localized
at the ends of the topological superconducting wire \cite{LutchynPRL10,1DwiresOreg, Mourik2012}.
%see also~\footnote{Another potential realization of a topological superconductor is
%Majorana quantum wire with two Majorana modes $\gamma_1$ and $\gamma_2$ localized at the
%opposite ends\cite{LutchynPRL10,1DwiresOreg, Mourik2012}}.
We assume here that the superconducting gap $\Delta$ is large, and develop an effective low-energy theory for the system valid at $E\ll \Delta$:
\begin{equation}
H=H_{\rm Leads} +H_{\rm Dot}+H_{\rm L-D}+ i\lambda (d_{\uparrow}+d_{\uparrow}^{\dagger})\gamma_1+ i\delta\gamma_1 \gamma_2.
\label{eq:Hamiltonian}
\end{equation}
Here
$H_{\rm Leads} = \sum_{\alpha=L,R}\sum_{k,\sigma} \epsilon_{k} c_{k\sigma,\alpha}^{\dagger}c_{k\sigma,\alpha}$,
$H_{\rm QD}=\sum_{\sigma}\epsilon_d d_{\sigma}^{\dagger}d_{\sigma}+U n_{\uparrow} n_{\downarrow}$,
and $H_{\rm L-D}=\sum_{\alpha=L,R}\sum_{k,\sigma} ( t_{\alpha k}c_{k\sigma,\alpha}^{\dagger} d_{\sigma}+h.c.)$
describe the leads, QD, and the Lead-QD coupling, respectively.
The operators $c_{k\sigma,\alpha}^{\dagger}$ ($d_{\sigma}^{\dagger}$) create a spin-$\sigma$ electron
in the $\alpha$-lead (the dot), $n_{\sigma}=d_{\sigma}^{\dagger}d_{\sigma}$,
$\epsilon_d$ is the chemical potential of the QD, $U$ is the QD on-site Coulomb interaction,
and $t_{\alpha k}$ ($\lambda$) is the tunneling coupling between the leads (TSC) and the QD. The splitting energy $\delta$ represents the finite overlap between two MZMs.
We note that the time-reversal symmetry is broken by the magnetic insulator
which also determines the spin polarization of the MZM. Without any loss of generality
we assume $\gamma_1$ only couples to the spin up channel of the QD.
The lead and QD Hamiltonians remain $\mathrm{SU}(2)$-invariant under spin rotation.
%We note that time-reversal symmetry is broken by the coupling to the topological superconductor, i.e. lead and QD Hamiltonians are $SU(2)$ invariant under spin-rotation.  Therefore, without any loss of generality one can assume MZM $\gamma_1$ only couples to spin up channel of the QD.

We first integrate out Majorana operators $\gamma_1$ and $\gamma_2$, which leads to the self-energy $\Sigma(\omega)$ (defined below). We assume that the QD is in the single-occupancy regime $U\gg |\epsilon_d|\gg \lambda,\Gamma$ with $\Gamma$ being the broadening of the QD level due to normal leads $\Gamma=\Gamma_L+\Gamma_R$ with
$\Gamma_{\alpha}=\pi|t_{\alpha}|^2 \rho_F$; here $\rho_F$ is the density state of the leads at the Fermi level. In this limit, one can use a
slave boson approximation for an infinite-U Anderson model\cite{ColemanPRB83,LeeRMP06} where the double occupancy of the QD is suppressed. Following standard procedure \cite{ColemanPRB83,LeeRMP06}, one can introduce the auxiliary boson $b$ and fermion $f_{\sigma}$ in order to $d_{\sigma}\rightarrow f_{\sigma}b^{\dagger}$, with the constraint $b^{\dagger}b+\sum_{\sigma}f_{\sigma}^{\dagger}f_{\sigma}=1$.
Within the slave boson mean field approximation (SBMF), we replace the bosonic field and the Lagrangian multiplier $\eta$ by their expectation values.
The mean field parameter $b$ and $\eta$ can be determined self-consistently by minimizing the free energy.
%In the calculation, we assume the $eV\ll T_K$ and thus neglect the dependence of the $eV$ in the SMBF calculations.
The detail of SBMF calculation in the presence of a MZM can be found in Ref. \cite{ChengPRX2014} (also see \cite{supp}). The SBMF approach decouples the spin-up channel from the spin-down channel and allows one to compute various correlation functions.

{\em Shot noise calculation}. We now use the Keldysh formalism~\cite{KamenevRev} to study current fluctuations.
Since two spin channels are decoupled within SBMF approximation, we drop the index $\sigma$ in this derivation. Given that MZM coupling breaks particle number conservation, the QD Green's function now acquires an anomalous contribution (e.g. $i\langle T_{c} d(t) d(t') \rangle$), and we need to work in the Nambu space $\mathbb{N}$.
We introduce a lead-QD basis
$\vec{\Psi}^{\dagger}=(\{c_{Lk}^{\dagger}, c_{Lk}\},d^{\dagger},d ,\{c_{Rk}^{\dagger}, c_{Rk}\})/\sqrt{2}$,
and write the action in this new space $\mathbb{S}$. The effective action can be written in terms of the full Green function $\breve{Q}$
\begin{eqnarray}
  S &=& S_{0}+ S_{L-D}+S_{\rm source},\\
 S_0+S_{L-D}&=&\int_C \int_C dt dt' \vec{\Psi}^{\dagger}(t) \breve{Q}^{-1}(t,t') \vec{\Psi}(t'),\\
 \breve{Q}_{kk'} &=&
 \begin{pmatrix}
  Q_{Lk,Lk'} & Q_{Lk,d} & Q_{Lk,Rk'} \\
  Q_{d,Lk'} & Q_{d,d} & Q_{d,Rk'} \\
  Q_{Rk,Lk'} & Q_{Rk,d} & Q_{Rk,Rk'}
 \end{pmatrix}
\end{eqnarray}
All matrix elements above have the same structure, e.g. $Q_{d,d}=\{\{G_{d\bar{d}},F_{dd}\},\{ F_{\bar{d}\bar{d}} , G_{\bar{d}d}\}\}$, see appendix \cite{supp} for details of $ \breve{Q}_{kk'}$.
After restoring the spin index, the retarded Green's functions are given by
\begin{eqnarray}
 G_{d\bar{d},\sigma}^{R}(\omega) &=& \frac{\omega+\widetilde{\epsilon}_{d}+i\widetilde{\Gamma}- \Sigma_{\sigma}(\omega)}
                       {(\omega+i\widetilde{\Gamma}-2\Sigma_{\sigma}(\omega))(\omega+i\widetilde{\Gamma})-\widetilde{\epsilon}_{d}^2},\\
 F_{dd,\sigma}^{R}(\omega) &=& \frac{-\Sigma_{\sigma}(\omega)}
                       {(\omega+i\widetilde{\Gamma}-2\Sigma_{\sigma}(\omega))(\omega+i\widetilde{\Gamma})-\widetilde{\epsilon}_{d}^2},
\end{eqnarray}
where $\Sigma_{\sigma}(\omega)=\lambda_{\sigma}^2 b^2\omega/(\omega^2-\delta^2)$ with $\lambda_{\uparrow}=\lambda b$ and $\lambda_{\downarrow}=0$. The effective broadening and energy of the QD level now read
$\widetilde{\Gamma}=\Gamma b^2$, $\widetilde{\epsilon}_{d}= \epsilon_{d}+\eta$.

We now consider current fluctuations through the left junction.  The current operator is given by
\begin{equation}
 I_{L}=\frac{i e}{\hbar} \sum_{k} \Big(  \widetilde{t}_{L k}c_{Lk}^{\dagger} d-  \widetilde{t}_{L k}^{*}d^{\dagger}c_{Lk}\Big)
      = \vec{\Psi}^{\dagger} \hat{M} \vec{\Psi},
\end{equation}
The 6-by-6 matrix $\hat{M}$ in $\mathbb{N}\otimes\mathbb{S}$ space for lead momentum $k$ is
$\hat{\rm M}_k=(ie/\hbar)\{ \{ 0 , M^{12}_k , 0 \},\{ M^{21}_k , 0 , 0 \},\{ 0,0,0 \} \}$
where $M^{21}_k=\bigl(\begin{smallmatrix}
-\widetilde{t}_{Lk}^{*}&0\\ 0 & -\widetilde{t}_{Lk}
\end{smallmatrix} \bigr)$
and $M^{12}_k=\bigl(\begin{smallmatrix}
\widetilde{t}_{Lk}&0\\ 0 & \widetilde{t}_{Lk}^{*}
\end{smallmatrix} \bigr)$.
Then, the action for the source term is
\begin{equation}
 S_{\rm source} = -\int_{C} dt A(t)I_L (t) = -\int_{-\infty}^{\infty} dt  \vec{\Psi}_a^{\dagger} \hat{A}_{ab}\hat{M} \vec{\Psi}_b .
\end{equation}
Here we rewrote the action in terms of the forward and backward components and performed Larkin-Ovchinnikov rotation\cite{KamenevRev}.
As a result, the source
$\hat{A}=A^{\alpha}\hat{\gamma}^{\alpha}$ is now a matrix in Keldysh $\mathbb{K}$ space, where $\alpha=cl,q$ with $\hat{\gamma}^{cl}=\mathbb{I}$ and
$\hat{\gamma}^{q}=\sigma_1$, see details in SI\cite{supp}. The generating function for this problem
$Z[A] = \int D[\{c_{Lk}^{\dagger} c_{Lk}\} d^{\dagger} d \{c_{Rk}^{\dagger} c_{Rk}\}] \;e^{i S}$
can be obtained in the following way\cite{KamenevRev} :
$\ln Z[A] = \rm{Tr} \ln \Big[ \breve{I} - \breve{Q} \hat{A}\hat{M}  \Big]$,
where the unit matrix $\breve{I}$ and the Green function $\breve{Q}$ are defined in $\mathbb{N}\otimes\mathbb{S}\otimes\mathbb{K}$ space,
$\hat{A}$ is in $\mathbb{K}$ space, and $\hat{M}$ is in  $\mathbb{N}\otimes\mathbb{S}$ space.
Finally, the symmetrized current noise for left junction can be written as
\begin{eqnarray}
 S_I(\omega, eV)&=&\int dt e^{i\omega t} \langle \delta I_L(t)\delta I_L(0)+\delta I_L(0)\delta I_L(t) \rangle\nonumber\\
               &=&  -\frac{1}{4} \frac{\delta^2 \ln Z[A]}{\delta A^q(\omega) \delta A^q(-\omega)} \bigg\vert_{A=0}
               \label{eq:shotnoise1}
\end{eqnarray}
where $\delta I_L(t)=I_L(t)-\langle I_L\rangle$, and an extra factor $1/2$ is to remove the doubling of the Hilbert space.
%After some manipulations, the zero-frequency expression for the shot noise becomes
%\begin{equation}\label{eq:shotnoise1}
% S_I(eV) = \frac{1}{4}\int \frac{d\omega}{2\pi} \sum_{kk'} \rm{Tr}
%             \Big\{ \breve{Q}_{kk'}\;\big(\hat{\gamma}^q\hat{M}_{k'}\big)\; \breve{Q}_{k'k}\;\big(\hat{\gamma}^q\hat{M}_{k}\big)   \Big\}
%\end{equation}
%Note that the matrix in $\mathbb{K}$ space is not written explicitly.
\begin{figure}
\centering
\includegraphics[width=3.2in,clip]{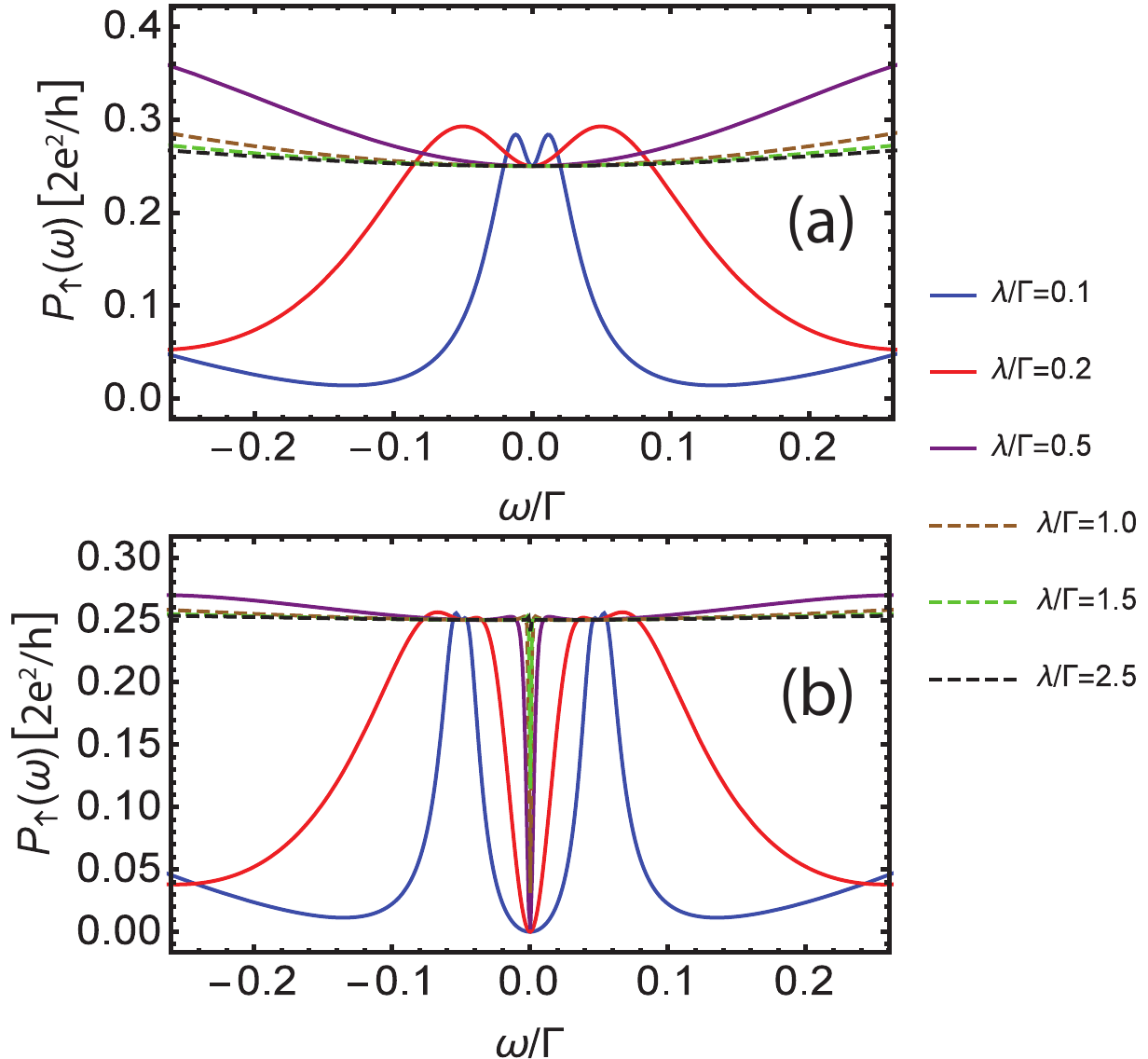}
\caption{ The power spectrum $P_{\uparrow}(\omega)$
for $\Gamma_L=\Gamma_R$ and different $\lambda$. Panel (a) no splitting of MZMs $\delta/\Gamma=0$.
The non-monotonic dependence of $P_{\uparrow}(\omega)$ originates from the non-trivial dependence of the P-H contribution $\mathbb{A}_{A}$ to the shot noise; (b) splitting energy $\delta/\Gamma=0.05$.
Here we used $\epsilon_d=0$.
}
\vspace{-0.1in}
\label{fig:Spectrum}
\end{figure}
The details of the evaluation of Eq.\eqref{eq:shotnoise1} are presented in SI\cite{supp}. At zero temperature, the shot noise is given by
\begin{equation}
S_I(eV)= \sum_{\sigma} \int\limits_{-eV/2}^{eV/2} d\omega P_{\sigma}(\omega),
 \label{eq:SNformula}
\end{equation}
where $P_{\sigma}(\omega)= (2 e^2/h)\Big( \mathbb{A}_{\rm N}^{\sigma}(\omega)+\mathbb{A}_{\rm A}^{\sigma}(\omega) \Big)$.
Here $P_{\sigma}(\omega)$ is the power spectrum of noise for each spin with $\mathbb{A}_{\rm N/A}^{\sigma}$ being the contributions to the noise from particle-particle (P-P)/particle-hole(P-H) channels, respectively. After tedious calculations (see SI\cite{supp} for details), one finds
\begin{eqnarray}
  \mathbb{A}_{\rm N}^{\sigma} &=& 2\widetilde{\Gamma}_L\widetilde{\Gamma}_R\big(|G_{d\bar{d},\sigma}^R|^2+|G_{\bar{d}d,\sigma}^R|^2\big)
                               +4\widetilde{\Gamma}_L^2|F_{dd,\sigma}^R|^2 \nonumber\\
                       &&        -8\widetilde{\Gamma}_L^2\widetilde{\Gamma}_R^2 \big(|G_{d\bar{d},\sigma}^R|^4+|G_{\bar{d}d,\sigma}^R|^4\big)
                               - 16\widetilde{\Gamma}_L^4|F_{dd,\sigma}^R|^4  \nonumber\\
                        &&     - 16\widetilde{\Gamma}_L^3\widetilde{\Gamma}_R\big(|G_{d\bar{d},\sigma}^R|^2+|G_{\bar{d}d,\sigma}^R|^2\big) |F_{dd,\sigma}^R|^2,\\
  \mathbb{A}_{\rm A}^{\sigma} &=& \widetilde{\Gamma}_L^2 \Big[ \big( F_{dd,\sigma}^R +F_{dd,\sigma}^A \big)^2  \nonumber\\
                        &&    -8(\widetilde{\Gamma}_L^2-\widetilde{\Gamma}_R^2)
                             \frac{|F_{dd,\sigma}^R|^2}{\Sigma_{\sigma}} \big( F_{dd,\sigma}^R +F_{dd,\sigma}^A \big) \nonumber\\
                        && +16(\widetilde{\Gamma}_L-\widetilde{\Gamma}_R)^2((\widetilde{\Gamma}_L+\widetilde{\Gamma}_R)^2+\widetilde{\epsilon}_d^2)
                           \frac{|F_{dd,\sigma}^R|^4}{\Sigma_{\sigma}^2}\Big],
\end{eqnarray}
where $G_{\bar{d}d}^R$ can be obtained from $G_{d\bar{d}}^R$ by $\epsilon_d\rightarrow -\epsilon_d$.
The P-H contribution is vanishing at zero frequency $\mathbb{A}_{\rm A}^{\sigma}(\omega) \sim \omega^2$.
Here we assume a symmetric bias $V_L=-V_R$.
%with $T_{\rm N}^{\sigma}(\omega)$ and $T_{\rm A}^{\sigma}(\omega)$ being the transmission probabilities.

{\em Results and Discussions}. Before presenting the results for an interacting QD problem, it is instructive to consider first a non-interacting spinless model, for which the results can be easily obtained by setting $\eta=0$ and $b=1$ in $P_{\uparrow}(\omega)$\eqref{eq:SNformula}. The power spectrum $P(0)$ at $T=0$  and $\delta=0$ is
\begin{align}
 P_{\lambda\neq 0}(0)& = \frac{2e^2}{h} \frac{\Gamma_L \Gamma_R}{\Gamma^2}=\left.\frac{e^2}{2h}\right|_{\Gamma_L=\Gamma_R}\\
 P_{\lambda=0}(0)& \!=\! \frac{2e^2}{h} \frac{4\Gamma_L \Gamma_R}{\Gamma^2\!+\!\epsilon_d^2}\left(1\!-\! \frac{4\Gamma_L \Gamma_R}{\Gamma^2\!+\!\epsilon_d^2}\right)\!=\!\left. \frac{2e^2}{h} \frac{\Gamma^2 \epsilon_d^2}{(\Gamma^2\!+\!\epsilon_d^2)^2} \right|_{\Gamma_L=\Gamma_R}
     \label{eq:SN_RL_MF}
\end{align}
One can see that coupling to MZM dramatically modifies the shot noise. For example, at the symmetric point the shot noise power does not depend on
$\epsilon_d$ and is given by $e^2/2h$ whereas without MZM $P_{\lambda=0}(0)$ depends on $\epsilon_d$
and is zero on resonance $\epsilon_d=0$. By tuning the coupling asymmetry $\Gamma_L/\Gamma_R$ or QD energy level $\epsilon_d$,
one should observe a qualitative different behaviour for the cases with and without MZMs, see appendix \cite{supp} for more details.
%in the former case, both tunneling conductance and shot noise are increasing and reaching the maximum at $\Gamma_L=\Gamma_R$
%whereas in the latter case the increase of the conductance is accompanied with decreasing of $P(0)$.
The shot noise at finite bias $eV$ is given by Eq.\eqref{eq:SNformula}.
In order to understand the $eV\neq 0$ results, we plot the power spectrum $P(\omega)$ in Fig. \ref{fig:Spectrum}~(a),
which shows a two-peak structure.
For $\lambda\ll\Gamma$, we find that the width between the two peaks $\sim\lambda^2/\Gamma$, and for $\lambda\gg\Gamma$,
this width becomes $\sim\Gamma$.
%Therefore, in small $\lambda$ limit, the bias $eV$ must be tuned to a very small value in order to observe the $1/2$ value.

\begin{figure}
\centering
\includegraphics[width=2.5in,clip]{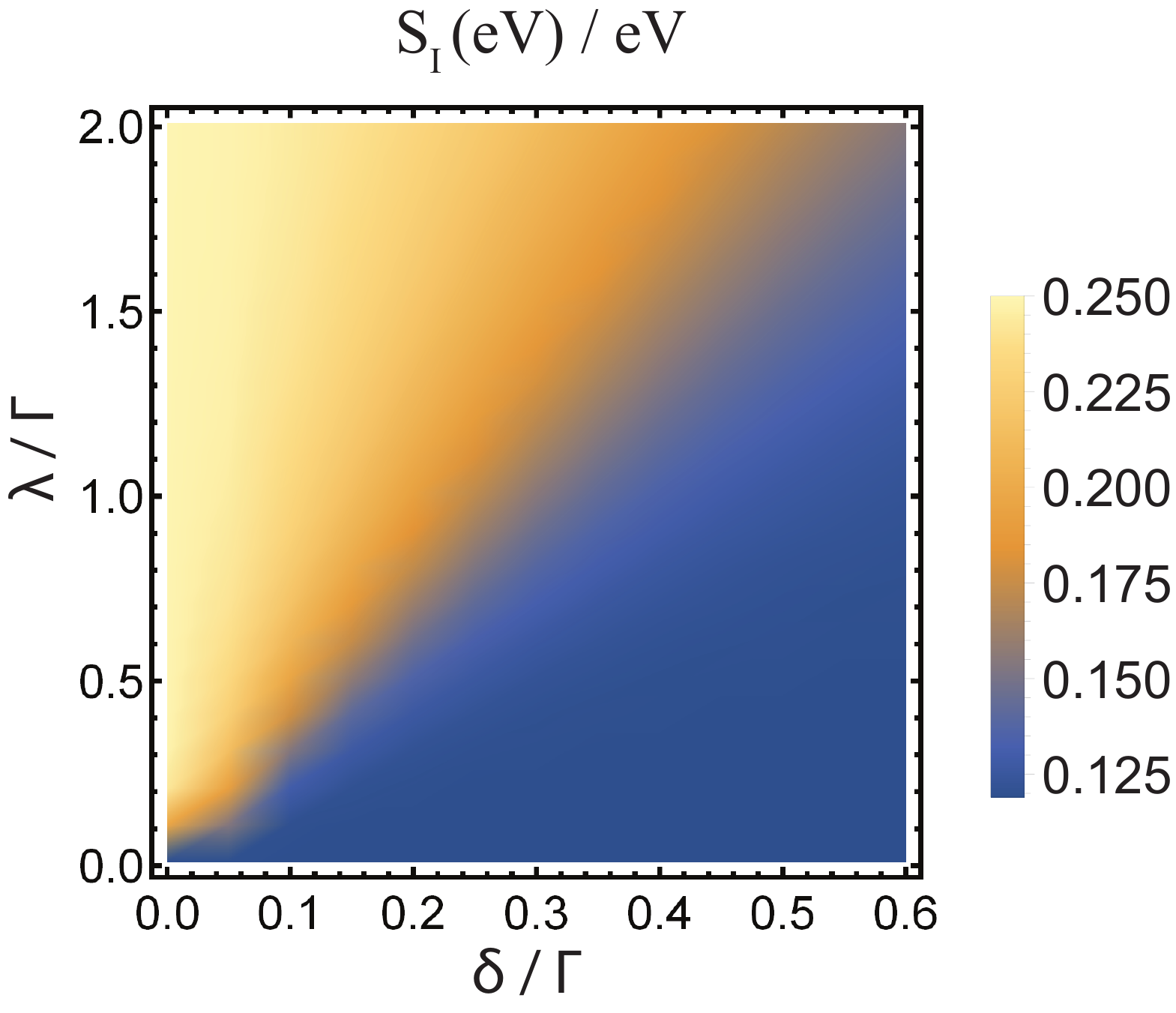}
\caption{Spinless non-interacting QD: the dependence of the shot noise $S_I(eV)/eV$ (measured in units of $2e^2/h$) at finite bias $eV/\Gamma=0.1$ on $\lambda$ and $\delta$. Here $\Gamma_L=\Gamma_R$, $\epsilon_d/\Gamma=-0.4$.
}
\vspace{-0.15in}
\label{fig:ShotNvsLamDel}
\end{figure}

We now discuss results at finite splitting $\delta \neq 0$. As shown in Fig. \ref{fig:Spectrum}~(b), the spectral function for a finite $\delta$ exhibits two peaks at small $\omega$. When $\lambda\ll \Gamma$, the position of the peak is at
$\pm \delta$. Thus, in order to observe the predicted value $P(0)=e^2/2h$, one should adjust the voltage to be $\lambda^2/\Gamma\gg eV\gg\delta$.
When $\lambda\gg\Gamma$, the width of the splitting is $\Gamma\delta^2/\lambda^2$. Thus,
the condition to observe $P(0)=e^2/2h$ value is $\Gamma\gg eV\gg \Gamma\delta^2/\lambda^2$.
We plot the shot noise as both a function of the $\lambda$ and $\delta$ in Fig. \ref{fig:ShotNvsLamDel}. One can see that the larger the splitting energy
$\delta$, the larger Majorana coupling $\lambda$ is needed to observe the predicted value for the shot noise $P(0)=e^2/2h$.
The effect of varying $\epsilon_d$ is discussed in \cite{supp}.

The conclusion based on the results of the spinless non-interacting problem is that Majorana 
coupling qualitatively modifies the shot noise through the QD. Thus, the combination of the 
conductance and shot noise measurements allow one to clarify the nature of the zero-bias conduction 
feature, see Table~\ref{tableI}. Even though the spinful problem is more complicated,
we show that this qualitative feature persists in the presence of interactions and allows one to distinguish between the 
Majorana and Kondo origin of the zero-bias feature in the tunneling conductance. 
We now consider the QD in the limit of single-occupancy $U\gg |\epsilon_d|\gg\Gamma, \lambda$ and $eV\ll T_K$ with $T_K$ being the Kondo temperature. We first analyze the case of no splitting $\delta=0$. A recent study based on SBMF approach~\cite{ChengPRX2014} shows that a crossover from Kondo- and Majorana-dominated regimes can be realized by tuning the coupling $\lambda$ .
For $\lambda\ll \lambda_{c} \equiv \sqrt{T_k/\Gamma} |\epsilon_d|$,
Kondo effect is important \cite{ChengPRX2014}: the renormalized coupling corresponds to
Kondo temperature $\tilde {\Gamma}\equiv\Gamma b^2=T_K=\Lambda \exp(-\pi|\epsilon_d|/2\Gamma)$ and the renormalized energy level is $\tilde{\epsilon}_d\equiv|\epsilon_d+\eta|\sim \Gamma b^4$ (Here $\Lambda$ is the bandwidth and $b \ll 1$ is the variational parameter).
When $\lambda \gg \lambda_c$, the parameter $b\sim \lambda/|\epsilon_d|$ is determined by the Majorana coupling rather than the Kondo temperature. One can see that in the perturbative regime $|\epsilon_d|\gg\Gamma, \lambda$ corresponding to $b\ll 1$, the position of the renormalized level is close to the Fermi energy $\widetilde{\epsilon}_{d}\sim \Gamma b^4\ll \widetilde{\Gamma}$ \cite{ChengPRX2014}.
In both cases the spin-down channel shows perfect transmission (i.e. linear conductance $G=e^2/h$), and, thus,  its contribution
to the shot noise is zero. On the other hand, the shot noise for the spin-up channel, due to the coupling to MZM, corresponds to
the universal
value $e^2/2h$ independent of $\epsilon_d$. The conductance and shot noise for spinful QD can be summarized as follows.
The linear conductance for $|\epsilon_d|\gg\lambda,\Gamma$ reads
\begin{equation}
 \left. G \right |_{\Gamma_L\!=\!\Gamma_R} =
    \frac{e^2}{h} \left(\frac{1}{2}+1\right)=\frac{3e^2}{2h} ,
  \label{eq:G_compare}
\end{equation}
which is consistent with the numerical renormalization group calculation~\cite{LeePRB13}.
The shot noise power is
\begin{equation}
 \left. P(0)\right|_{\Gamma_L\!=\!\Gamma_R} =
    \frac{2e^2}{h} \left( \frac{1}{4} + 0\right)=\frac{e^2}{2h}.
  \label{eq:SN_compare}
\end{equation}
The results beyond the $|\epsilon_d|\gg\lambda, \Gamma$ limit can be obtained numerically and are discussed in \cite{supp}.

\begin{figure}
    \includegraphics[width=3.0in,clip]{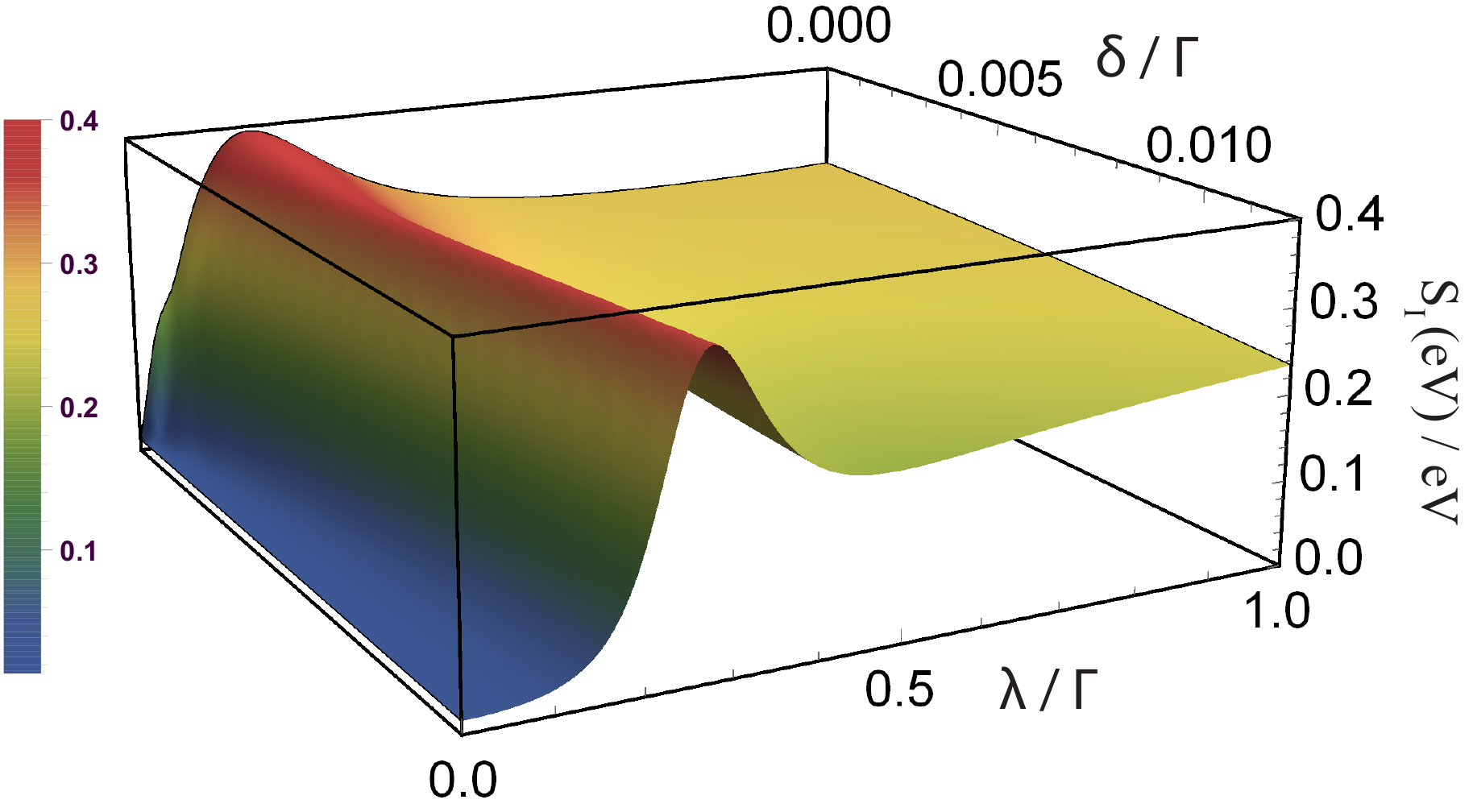}
\caption{ Spinful QD in the single-occupancy regime:
The dependence of the shot noise $S_I(eV)/eV$ (measured in units of $2e^2/h$) as a function of $\lambda$ and $\delta$.
Here $\epsilon_d/\Gamma=-10.0$, $\Gamma_L=\Gamma_R$, $eV/\Gamma=0.001$, and $\Lambda/\Gamma=30.0$.
The non-monotonic behavior as a function of $\lambda$ originates from the P-H contribution $\mathbb{A}_{A}$.
The contribution to $S_I(eV)$ from the spin-down channel is negligibly small in this parameter regime.
    }
    \vspace{-0.1in}
    \label{fig:SBMFT_SN}
\end{figure}

We now consider the effect of a finite energy splitting $\delta\neq 0$ and a finite bias $eV\neq 0$
which is important for the experimental detection of the effect we predict here.
The shot noise $S_I(eV)/eV$ as a function of $\lambda$ and $\delta$ is shown in the Fig. \ref{fig:SBMFT_SN}.
One can see that in order to resolve the quantized value $P(0)=e^2/2h$, one has to satisfy the following
conditions: a) in the regime $b\lambda\ll b^2\Gamma$, the voltage should be $\lambda^2/\Gamma \gg eV \gg \delta$;
b) in the $\lambda\gg b\Gamma$ regime, the condition becomes $b^2 \Gamma \gg eV \gg \Gamma \delta^2/\lambda^2$.
It is thus clear that in the Majorana-dominated regime, i.e. $\lambda\gg b|\epsilon_d|\gg b\Gamma$,
the voltage should satisfy condition b), in which case the shot noise power spectrum exhibits a plateau
around $S_I(eV)/eV=e^2/2h$, see Fig. \ref{fig:SBMFT_SN}.
One  can also notice that the width of the plateau around $S_I(eV)/eV=e^2/2h$
gradually shrinks with increasing $\delta$. In the limit $\lambda\gtrsim |\epsilon_d|$, the renormalized energy level $\widetilde{\epsilon}_d$ shifts away from
the Fermi level since $b\sim 1$, which, in turn, suppresses the conductance at zero bias and enhances the shot noise, see discussion in appendix \cite{supp}.

%\DEL{Note that Majorana zero modes also emerge in certain quantum impurity models ~\cite{Emery&Kivelson,Affleck93TCK,Affleck95TIK,Schiller&Hershfield}.
%i.e. two channel Kondo models \cite{Emery&Kivelson,Affleck93TCK} and two impurity Kondo models \cite{Affleck95TIK}.
%For example, the half-quantized value of the conductance was also predicted in a double quantum dot model at
%the two channel quantum critical point \cite{Zarand06TCK}. However, this requires fine-tuning in contrast to 
%the Majorana physics discussed here, which is a robust phenomenon protected by the gap of a topological superconductor. \cite{GGTCK,Mebrahtu13}.
%}

%{\it Conclusions.} We study current fluctuations in a QD coupled to a topological superconductor, and show that Majorana modes qualitatively modify the result for the shot noise. We argue that the combination of the average current measurement and its fluctuations
%allows one to distinguish
%between topological (Majorana) and non-topological (e.g., Kondo) origin of the zero-bias anomaly in the QDs. Thus, our results have important implications for the tunneling experiment trying to detect Majorana zero modes.

R.L. acknowledges the hospitality of the Aspen Center for Physics supported by NSF grant \#1066293, where
part of this work was done.

\appendix

\begin{widetext}
\section{Appendix for ``Probing Majorana Physics in Quantum Dot Shot Noise Experiments''}

In this appendix we will provide
1) the details of the derivation of the shot noise formula $S_I(eV)$ defined in Eq.(10) of the main text;
2) the discussion of the effects of the dot energy level $\epsilon_d$;
3) the main steps in slave boson mean field approach;
4) the analysis of false-positive signatures.

\subsection{The derivation of the shot noise $S_I(eV)$ in Eq.(10) of the main text}

We first derive the full impurity Green function $Q_{d,d}$:
\begin{equation}
   Q_{d,d}(t-t')=
 \begin{pmatrix}
  G_{d\bar{d}} & F_{dd}  \\
  F_{\bar{d}\bar{d}} & G_{\bar{d}d}
 \end{pmatrix}\!=\!-\!
 \begin{pmatrix}
  i\langle T_c d(t) d^{\dagger}(t') \rangle & i\langle T_c d(t) d(t')\rangle  \\
  i\langle T_c d^{\dagger}(t) d^{\dagger}(t')\rangle & i\langle T_c d^{\dagger}(t) d(t')\rangle
 \end{pmatrix}.
\end{equation}
For the clarity of presentation, we drop the tilde index in $\widetilde{\Gamma}$ and $\widetilde{\epsilon}_d$. The dependence on $b$ and $\eta$ can be easily restored. For simplicity, we drop the spin index $\sigma$ in this derivation.
Using the following convention for the Nambu spinors: $\vec{\Psi}_{k\alpha}^{\dagger}=(c_{\alpha k}^{\dagger}, c_{\alpha k})/\sqrt{2}$
and $\vec{\Psi}_{d}^{\dagger}=(d^{\dagger}, d)/\sqrt{2}$, where $\alpha=L, R$ is the lead index, the effective Keldysh action
for each spin channel now reads
\begin{equation}
  S = S_{0}+ S_{L-D},
\end{equation}
where
\begin{eqnarray}
S_{0} &=& \sum_{kk',\alpha} \int_{C}\int_{C} dtdt' \vec{\Psi}_{k\alpha}^{\dagger}(t)\,
               \breve{Q}_{0,kk'\alpha}^{-1}(t,t') \,\vec{\Psi}_{k\alpha}(t')\nonumber\\
       && + \int_{C}\int_{C} dtdt' \vec{\Psi}_{d}^{\dagger}(t)\, \breve{Q}_{0,dd}^{-1}(t,t')\, \vec{\Psi}_{d}(t'),\\
S_{L-D} &=& -\sum_{k\alpha}\int_{C} dt \Big( t_{\alpha k}c_{\alpha k}^{\dagger} d+c.c.  \Big)\\
       &=&  -\sum_{k\alpha}\int_{C} dt (\vec{\Psi}_{k\alpha}^{\dagger}(t) M_{T,\alpha k} \vec{\Psi}_{d}(t)+h.c.),
\end{eqnarray}
are the actions for leads, QD, and Lead-QD couling, and
$M_{T,\alpha k}=\bigl(\begin{smallmatrix}
t_{\alpha k}&0\\ 0 & -t_{\alpha k}^{*}\end{smallmatrix} \bigr)$. Here the integration is over the Keldysh contour. The free lead Green's
function $\breve{Q}_{0,kk'\alpha}$ and free QD Green's function $\breve{Q}_{0,dd}$ (with MZM coupling)
in the Nambu space $\mathbb{N}$ can be written as
\begin{eqnarray}
 Q_{0,kk'\alpha}(t-t') = \delta_{kk'}
 \begin{pmatrix}
  g_{\alpha k}^0(t-t') &0  \\
  0 & \widetilde{g}_{\alpha k}^0(t-t')
 \end{pmatrix},
  Q_{0,dd}(t-t') =
 \begin{pmatrix}
  G_{0,d\bar{d}}(t-t') & F_{0,dd}(t-t')   \\
  F_{0,\bar{d}\bar{d}}(t-t')  & G_{0,\bar{d},d}(t-t')
 \end{pmatrix},
\end{eqnarray}
where $\widetilde{g}_{\alpha k}^0(t-t')$ ($G_{0,\bar{d}d}(t-t')$) is the P-H conjugation of
$g_{\alpha k}^0(t-t')$ ($G_{0,d\bar{d}}(t-t')$).
We first perform Larkin-Ovchinnikov
rotation~\cite{KamenevRev}.
The retarded, advanced and Keldysh components of the Green's function are defined as
\begin{align}
G_{ab}(t,t')=-i\langle \Psi_a(t) \Psi^\dag_b(t') \rangle=\left(
                                                      \begin{array}{cc}
                                                        G^R(t,t') & G^K(t,t') \\
                                                        0 & G^A(t,t') \\
                                                      \end{array}
                                                    \right)
\end{align}
We remind the reader that the relationship between Keldysh Green's function before and after LO rotation is
\begin{align}
\left(
                                                      \begin{array}{cc}
                                                        G^R(t,t') & G^K(t,t') \\
                                                        0 & G^A(t,t') \\
                                                      \end{array}
                                                    \right)=\frac 1 2 \left(
                                                      \begin{array}{cc}
                                                        1 & 1 \\
                                                        1 & -1 \\
                                                      \end{array}
                                                    \right) \left(
                                                      \begin{array}{cc}
                                                        G^T(t,t') & G^<(t,t') \\
                                                        G^>(t,t') & G^{\tilde T}(t,t') \\
                                                      \end{array}\right) \left(
                                                      \begin{array}{cc}
                                                        1 & 1 \\
                                                        -1 & 1 \\
                                                      \end{array}\right)
\end{align}
After performing the Gaussian integration and Fourier transformation, one can obtain the full impurity Green function
\begin{equation}
 Q_{dd}(\omega)^{-1} = Q_{0,dd}(\omega)^{-1} -
 \begin{pmatrix}
  \sum_{\alpha,k}|t_{\alpha k}|^2 g_{\alpha k}^0(\omega) & 0\\
  0 & \sum_{\alpha,k}|t_{\alpha k}|^2 \widetilde{g}_{\alpha k}^0(\omega)
 \end{pmatrix}.\label{eq:QddGreen}
\end{equation}
The free QD Green function (with MZM coupling) can be written as
\begin{equation}
 Q_{0,dd}(\omega) =
 \begin{pmatrix}
  G_{0,d\bar{d}}^R & G_{0,d\bar{d}}^K & F_{0,dd}^R & F_{0,dd}^K \\
  0 & G_{0,d\bar{d}}^A & 0 & F_{0,dd}^A \\
   F_{0,\bar{d}\bar{d}}^R & F_{0,\bar{d}\bar{d}}^K & G_{0,\bar{d}d}^R & G_{0,\bar{d}d}^K \\
  0 & F_{0,\bar{d}\bar{d}}^A & 0 & G_{0,\bar{d}d}^A \\
 \end{pmatrix},
\end{equation}
where
\begin{eqnarray}
 G_{0,d\bar{d}}^R(\omega) &=& [G_{0,d\bar{d}}^A(\omega)]^{*}=\frac{\omega+i\eta+\epsilon_d-\Sigma(\omega)}{(\omega+i\eta-2\Sigma(\omega))(\omega+i\eta)-\epsilon_d^2}\\
 G_{0,\bar{d}d}^R(\omega) &=& [G_{0,\bar{d}d}^A(\omega)]^{*}=\frac{\omega+i\eta-\epsilon_d-\Sigma(\omega)}{(\omega+i\eta-2\Sigma(\omega))(\omega+i\eta)-\epsilon_d^2}\\
F_{0,dd}^R(\omega) &=& F_{0,\bar{d}\bar{d}}^R(\omega)= [F_{0,dd}^A(\omega)]^{*}=[F_{0,\bar{d}\bar{d}}^A(\omega)]^{*}=\frac{-\Sigma(\omega)}{(\omega+i\eta-2\Sigma(\omega))(\omega+i\eta)-\epsilon_d^2}.
 \end{eqnarray}
Since we assume that $E \ll \Delta$, all Keldysh components
($G_{0,d\bar{d}}^K$, $G_{0,\bar{d}d}^K$, $F_{0,dd}^K$, $F_{0,\bar{d}\bar{d}}^K$) are zero.
The Green functions of the free lead electron are related to tunneling rate
$\Gamma_{\alpha}=\pi |t_{\alpha}|^2 \rho_{F}$ ($\rho_F$ is the density of state of the leads near Fermi level)
\begin{eqnarray}
 \sum_{k}|t_{\alpha k}|^2 g_{\alpha k}^0(\omega) &=&
\begin{pmatrix}
 -i\Gamma_{\alpha} & -2i \Gamma_{\alpha} (1-2 n_{\alpha})\\
 0 & i\Gamma_{\alpha}
\end{pmatrix},\label{eq:gPGreen}\\
 \sum_{k}|t_{\alpha k}|^2 \widetilde{g}_{\alpha k}^0(\omega) &=&
\begin{pmatrix}
 -i\Gamma_{\alpha} & -2i \Gamma_{\alpha} (1-2 \widetilde{n}_{\alpha})\\
 0 & i\Gamma_{\alpha}
\end{pmatrix}.\label{eq:gHGreen}
\end{eqnarray}
One notices that $g_{\alpha k}^{0,R}(\omega)=-\widetilde{g}_{\alpha k}^{0,A}(-\omega)$
and $g_{\alpha k}^{0,K}(\omega)=-\widetilde{g}_{\alpha k}^{0,K}(-\omega)$.
Here, $n_{\alpha}$ is the fermi distribution function of the $\alpha$ lead with
chemical potential $\mu_{\alpha}$, and $\widetilde{n}_{\alpha}$ corresponds
to the fermi distribution function with $-\mu_{\alpha}$.
We consider a symmetric source-drain bias ($\mu_L=eV/2$ and $\mu_R=-eV/2$), and thus have
$\widetilde{n}_L=n_R$ and $\widetilde{n}_R=n_L$. We note in passing here that for asymmetric couplings $\Gamma_R \neq \Gamma_L$, the steady-state distribution function for the dot depends on Majorana coupling $\lambda$.

%In addition, in the presence of MZM coupling, some relations, e.g. $G_{d\bar{d}}^K=(\Gamma_L(1-2n_L)+\Gamma_R(1-2n_R))(G_{d\bar{d}}^R-G_{d\bar{d}}^A)/(\Gamma_L+\Gamma_R)$,
%are no longer true.

To derive the shot noise, we then consider Eq. (7) of the main text. The source action on the Keldysh contour reads
\begin{align}
 S_{\rm source} = -\int_{C} dt A(t)I_L (t) = -\int_{-\infty}^{\infty} dt  \left[\vec{\Psi}_{+}^{\dagger} A_+(t) \hat{M} \vec{\Psi}_+-\vec{\Psi}_{-}^{\dagger}A_{-}(t) \hat{M} \vec{\Psi}_-\right].
\end{align}
where $\vec{\Psi}_{\pm}^{\dagger}$ and $A_{\pm}(t)$ are the fermionic fields and source fields on the forward
and backward branches of the Keldysh contour. We again perform Larkin-Ovchinnikov
rotation~\cite{KamenevRev}: $\psi_{1,2}=(\psi_+\pm \psi_-)/\sqrt{2}$ and $\psi^\dag_{1,2}=(\psi_+\mp \psi_-)/\sqrt{2}$ and $A^{cl/q}=(A_+\pm A_-)/2$. Thus, the source term now becomes
\begin{align}
 S_{\rm source} = -\sum_{a,b=1,2}\int_{-\infty}^{\infty} dt  \vec{\Psi}_a^{\dagger} \hat{A}_{ab}\hat{M} \vec{\Psi}_b .
\end{align}
where $\hat{A}=A^{\alpha}\hat{\gamma}^{\alpha}$ is now a matrix in Keldysh $\mathbb{K}$ space,
where $\alpha=cl,q$ with $\hat{\gamma}^{cl}=\mathbb{I}$ and $\hat{\gamma}^{q}=\sigma_1$.
Using Eqs.(5), (10) and (11), one finds that the shot noise is given by
\begin{eqnarray}
 S_I(eV) &=& \frac{1}{4}\int \frac{d\omega}{2\pi} \sum_{kk'} \mathbf{Tr} \Big\{ \breve{Q}_{kk'}\;
           \big(\hat{\gamma}^q\hat{M}_{k'}\big)\; \breve{Q}_{k'k}\;\big(\hat{\gamma}^q\hat{M}_{k}\big)   \Big\} \nonumber\\
           &=& \frac{1}{4} \Big(\frac{ie}{\hbar}\Big)^2 \int\frac{d\omega}{2\pi} \sum_{kk'} \mathbf{Tr}
               \Big\{ Q_{Lk,d}\; (\hat{\gamma}^{q} M_{k'}^{21})\; Q_{Lk',d}\; (\hat{\gamma}^{q} M_{k}^{21})+
                  Q_{d,Lk'}\; (\hat{\gamma}^{q} M_{k'}^{12})\; Q_{d,Lk}\; (\hat{\gamma}^{q} M_{k}^{12}) \nonumber\\
           && \quad\quad\quad\quad\quad\quad + Q_{Lk,Lk'}\; (\hat{\gamma}^{q} M_{k'}^{12})\; Q_{d,d}\; (\hat{\gamma}^{q} M_{k}^{21}) +
              Q_{d,d}\; (\hat{\gamma}^{q} M_{k'}^{21})\; Q_{Lk',Lk}\; (\hat{\gamma}^{q} M_{k}^{12}) \Big\}
\end{eqnarray}
where $Q_{\alpha k,\alpha k'}(\omega)$, and $Q_{\alpha k,d}(\omega)$ are Fourier transform of
\begin{eqnarray}
 Q_{\alpha k,\alpha k'}(t-t') &=&
 \begin{pmatrix}
  G_{\alpha k,\bar{\alpha k'}} & F_{\alpha k,\alpha k'}  \\
  F_{\bar{\alpha k},\bar{\alpha k'}} & G_{\bar{\alpha k},\alpha k'}
 \end{pmatrix}=-
 \begin{pmatrix}
  i\langle T_c c_{\alpha k}(t) c_{\alpha k'}^{\dagger}(t') \rangle & i\langle T_c c_{\alpha k}(t) c_{\alpha k}(t')\rangle  \\
  i\langle T_c c_{\alpha k}^{\dagger}(t) c_{\alpha k}^{\dagger}(t')\rangle & i\langle T_c c_{\alpha k}^{\dagger}(t) c_{\alpha k'}(t')\rangle
 \end{pmatrix},\\
  Q_{\alpha k,d}(t-t') &=&
 \begin{pmatrix}
  G_{\alpha k,\bar{d}} & F_{\alpha k,d}  \\
  F_{\bar{\alpha k},\bar{d}} & G_{\bar{\alpha k},d}
 \end{pmatrix}=-
 \begin{pmatrix}
  i\langle T_c c_{\alpha k}(t) d^{\dagger}(t') \rangle & i\langle T_c c_{\alpha k}(t) d(t')\rangle  \\
  i\langle T_c c_{\alpha k}^{\dagger}(t) d^{\dagger}(t')\rangle & i\langle T_c c_{\alpha k}^{\dagger}(t) d(t')\rangle
 \end{pmatrix},
\end{eqnarray}
and one can define $Q_{d,\alpha k'}$ in a similar way.
We apply the matrix product in $\mathbb{S}$ space. For example, we expand the first term to find
\begin{eqnarray}
 &&\mathbf{Tr} \Big\{ Q_{Lk,d}\; (\hat{\gamma}^{q} M_{k'}^{21})\; Q_{Lk',d}\; (\hat{\gamma}^{q} M_{k}^{21}) \Big\} \\
     &=& \mathbf{Tr} \Big\{ t_{Lk}^{*} t_{Lk'}^{*} G_{Lk,\bar{d}}\; \hat{\gamma}^q\; G_{Lk',\bar{d}}\; \hat{\gamma}^q
           +t_{Lk} t_{Lk'} G_{\bar{Lk},d}\; \hat{\gamma}^q\; G_{\bar{Lk}',d}\; \hat{\gamma}^q
           +t_{Lk'} t_{Lk}^{*} F_{Lk,d}\; \hat{\gamma}^q\; F_{\bar{Lk}',\bar{d}}\; \hat{\gamma}^q
           +t_{Lk'}^{*} t_{Lk} F_{\bar{Lk},\bar{d}}\; \hat{\gamma}^q\; F_{Lk',d}\; \hat{\gamma}^q  \Big\}\nonumber
\end{eqnarray}
Similarly, we expand the three other terms
and obtain
\begin{equation}
 S_I(eV)=S_{I,N}(eV)+S_{I,A}(eV),
\end{equation}
where
\begin{eqnarray}
 S_{I,N}(eV)&=& -\frac{1}{4} \Big(\frac{e}{\hbar}\Big)^2 \int \frac{d\omega}{2\pi} \sum_{kk'}
     \mathbf{Tr} \Big\{ t_{Lk}^{*} t_{Lk'}^{*} G_{Lk,\bar{d}}\; \hat{\gamma}^q\; G_{Lk',\bar{d}}\; \hat{\gamma}^q
           +t_{Lk} t_{Lk'} G_{d,\bar{Lk}}\; \hat{\gamma}^q\; G_{d,\bar{Lk}'}\; \hat{\gamma}^q  \nonumber\\
        &&   \quad\quad\quad\quad
           -t_{Lk}^{*} t_{Lk'} G_{Lk,\bar{Lk}'}\; \hat{\gamma}^q\; G_{d,\bar{d}}\; \hat{\gamma}^q
           -t_{Lk'}^{*} t_{Lk} G_{d,\bar{d}}\; \hat{\gamma}^q\; G_{Lk',\bar{Lk}}\; \hat{\gamma}^q \Big\} + \text{P-H conjugation}
           \label{eq:SNregular}
\end{eqnarray}
and
\begin{eqnarray}
 S_{I,A}(eV)&=& -\frac{1}{2} \Big(\frac{e}{\hbar}\Big)^2 \int \frac{d\omega}{2\pi} \sum_{kk'}
     \mathbf{Tr} \Big\{- t_{Lk}^{*} t_{Lk'}^{*} F_{Lk,Lk'}\; \hat{\gamma}^q\; F_{\bar{d},\bar{d}}\; \hat{\gamma}^q
           -t_{Lk} t_{Lk'} F_{d,d}\; \hat{\gamma}^q\; F_{\bar{Lk},\bar{Lk}'}\; \hat{\gamma}^q  \nonumber\\
        &&   \quad\quad\quad\quad\quad\quad
           +t_{Lk}^{*} t_{Lk'} F_{Lk,d}\; \hat{\gamma}^q\; F_{\bar{Lk}',\bar{d}}\; \hat{\gamma}^q
           +t_{Lk}^{*} t_{Lk'} F_{d,Lk}\; \hat{\gamma}^q\; F_{\bar{d},\bar{Lk}'}\; \hat{\gamma}^q \Big\}.
           \label{eq:SNanomalous}
\end{eqnarray}
The first (i.e. normal) contribution also appears in the case without Majorana zero mode (particle-particle and hole-hole channels) whereas the second contribution represents an anomalous part due to the MZM coupling (particle-hole channel).

\begin{figure}[htp]
\centering
\includegraphics[width=4.2in,clip]{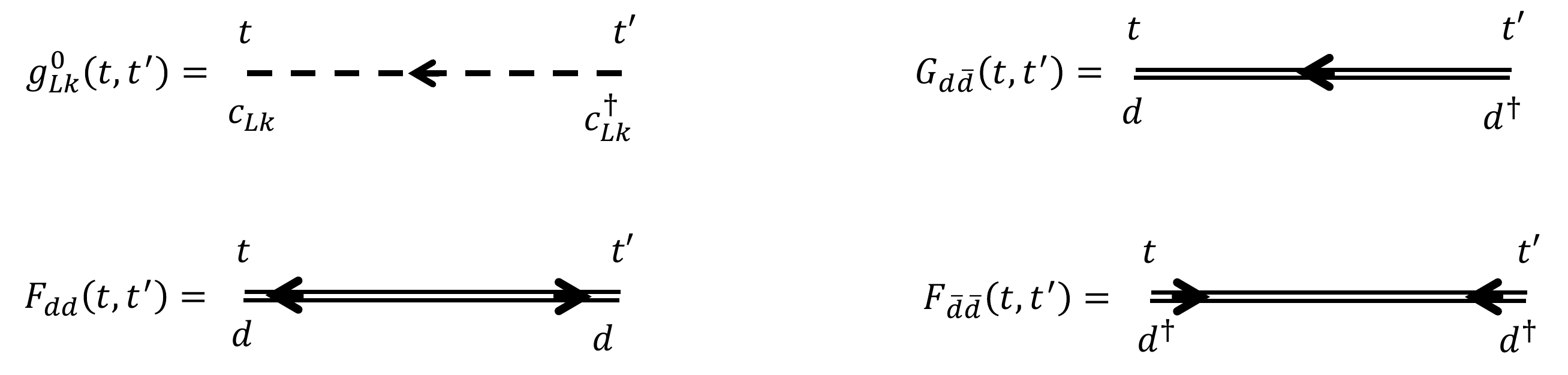}
\caption{ Diagrammatic representation of the free electron Green function $g_{Lk}^0(t,')$ and the full
impurity Green function $G_{d\bar{d}}(t,t')$, $F_{dd}(t,t')$ and $F_{\bar{d}\bar{d}}(t,t')$.
}
\vspace{0.2in}
\label{fig:NormalPropergator}
\end{figure}

\begin{figure}[htp]
\centering
\includegraphics[width=5.2in,clip]{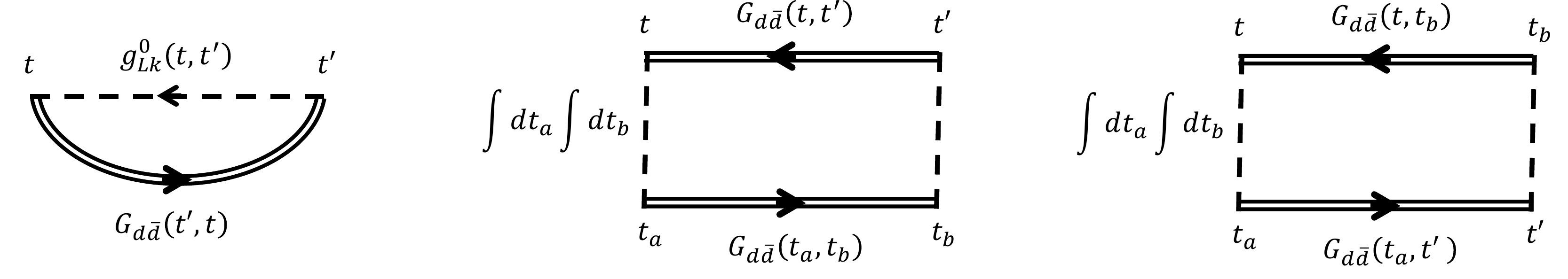}
\caption{ The normal contribution to the shot noise can be described by these three types of diagrams.
The left diagram corresponds to the first line in Eq. (\ref{eq:SN_regular1}), the middle diagram
(along with a similar diagram with arrow direction change)
corresponds to the second and third lines in Eq. (\ref{eq:SN_regular1}), and the right diagram
(along with a similar diagram with arrow direction change)
corresponds to the fourth and fifth lines in Eq. (\ref{eq:SN_regular1}).
}\vspace{0.2in}
\label{fig:SNDiaReg}
\end{figure}

We first consider the normal part. The Green function $ G_{Lk,\bar{Lk}'}$,  $G_{Lk,\bar{d}}$,
and $ G_{d,\bar{Lk}'}$ are related to the free electron Green function $g_{Lk}^0$ and the QD Green function
$G_{d,\bar{d}}$ via equations of motion, and thus can be written as
\begin{eqnarray}
 G_{Lk,\bar{Lk}'}(\omega)&=&g_{Lk}^0(\omega)\delta_{kk'}+t_{Lk} t_{Lk'}^{*}g_{Lk}^0(\omega) G_{d,\bar{d}}(\omega) g_{Lk'}^0(\omega),\\
 G_{Lk,\bar{d}}(\omega)&=& t_{Lk}g_{Lk}^0(\omega)G_{d,\bar{d}}(\omega), \\
 G_{d,\bar{Lk}}(\omega)&=& t_{Lk}^{*} G_{d,\bar{d}}(\omega) g_{Lk}^0(\omega).
\end{eqnarray}
We then insert those equations into Eq.(\ref{eq:SNregular}), and obtain
\begin{eqnarray}
  S_{I,N}(eV)&=& \frac{1}{4} \Big(\frac{e}{\hbar}\Big)^2 \int \frac{d\omega}{2\pi}
   \Bigg[ \mathbf{Tr}\Big\{ \sum_k |t_{Lk}|^2 g_{Lk}^0(\omega) \hat{\gamma}^q G_{d,\bar{d}}(\omega)\hat{\gamma}^q
   + G_{d,\bar{d}}(\omega)\hat{\gamma}^q \sum_k |t_{Lk}|^2 g_{Lk}^0(\omega) \hat{\gamma}^q \Big\} \nonumber\\
   &&\quad\quad\quad\quad\quad +  \mathbf{Tr} \Big\{ \sum_k |t_{Lk}|^2 g_{Lk}^0(\omega) G_{d,\bar{d}}(\omega) \sum_{k'} |t_{Lk'}|^2 g_{Lk'}^0(\omega)
            \hat{\gamma}^q G_{d,\bar{d}}(\omega) \hat{\gamma}^q  \nonumber\\
   &&\quad\quad\quad\quad\quad + G_{d,\bar{d}}(\omega) \hat{\gamma}^q \sum_{k'} |t_{Lk'}|^2 g_{Lk'}^0(\omega) G_{d,\bar{d}}(\omega) \sum_{k} |t_{Lk}|^2 g_{Lk}^0(\omega)
            \hat{\gamma}^q \nonumber\\
   &&\quad\quad\quad\quad\quad -   \sum_k |t_{Lk}|^2 g_{Lk}^0(\omega) G_{d,\bar{d}}(\omega)\hat{\gamma}^q
                       \sum_{k'} |t_{Lk'}|^2 g_{Lk'}^0(\omega) G_{d,\bar{d}}(\omega)\hat{\gamma}^q \nonumber\\
   &&\quad\quad\quad\quad\quad -  G_{d,\bar{d}}(\omega) \sum_k |t_{Lk}|^2 g_{Lk}^0(\omega) \hat{\gamma}^q
                       G_{d,\bar{d}}(\omega) \sum_{k'} |t_{Lk'}|^2 g_{Lk'}^0(\omega) \hat{\gamma}^q \Big\} \Bigg]+ \text{P-H conjugation}.
   \label{eq:SN_regular1}
\end{eqnarray}
In a diagrammatic representation defined in Fig. \ref{fig:NormalPropergator}, the normal contribution to the shot noise can be described by
the diagrams shown in Fig. \ref{fig:SNDiaReg}. We insert the Green functions $G_{d,\bar{d}}(\omega)$, $G_{\bar{d},d}(\omega)$,
$g_{\alpha k}^0(\omega)$, and $\widetilde{g}_{\alpha k}^0(\omega)$ from Eq. (\ref{eq:QddGreen}), (\ref{eq:gPGreen}), and (\ref{eq:gHGreen})
into Eq. (\ref{eq:SN_regular1}). We choose a symmetric source-drain bias, and take the zero
temperature limit. Then, we obtain the normal part of the shot noise for a spinless model, e.g. Eq. (10) and (11) of the main text:
\begin{equation}
 S_{I,N}(eV)=\frac{2 e^2}{h} \int\limits_{-eV/2}^{eV/2} \mathbb{A}_{N}(\omega) d\omega,
\end{equation}
where
\begin{eqnarray}
  \mathbb{A}_{\rm N}(\omega) &=& 2\Gamma_L\Gamma_R\big(|G_{d\bar{d}}^R(\omega)|^2+|G_{\bar{d}d}^R(\omega)|^2\big)
                               +4\Gamma_L^2|F_{dd}^R(\omega)|^2
                              -8\Gamma_L^2\Gamma_R^2 \big(|G_{d\bar{d}}^R(\omega)|^4+|G_{\bar{d}d}^R(\omega)|^4\big)\nonumber\\
                          &&     - 16\Gamma_L^4|F_{dd}^R(\omega)|^4
                            - 16\Gamma_L^3\Gamma_R\big(|G_{d\bar{d}}^R(\omega)|^2+|G_{\bar{d}d}^R(\omega)|^2\big) |F_{dd}^R(\omega)|^2,
\end{eqnarray}
Here, to derive this compact form, we use the relations among $G_{d\bar{d}}$, $G_{\bar{d}d}$, $F_{dd}$, and $F_{\bar{d}\bar{d}}$,
which can be obtained from Eq. (\ref{eq:QddGreen}).

\begin{figure}[t]
\centering
\includegraphics[width=4.5in,clip]{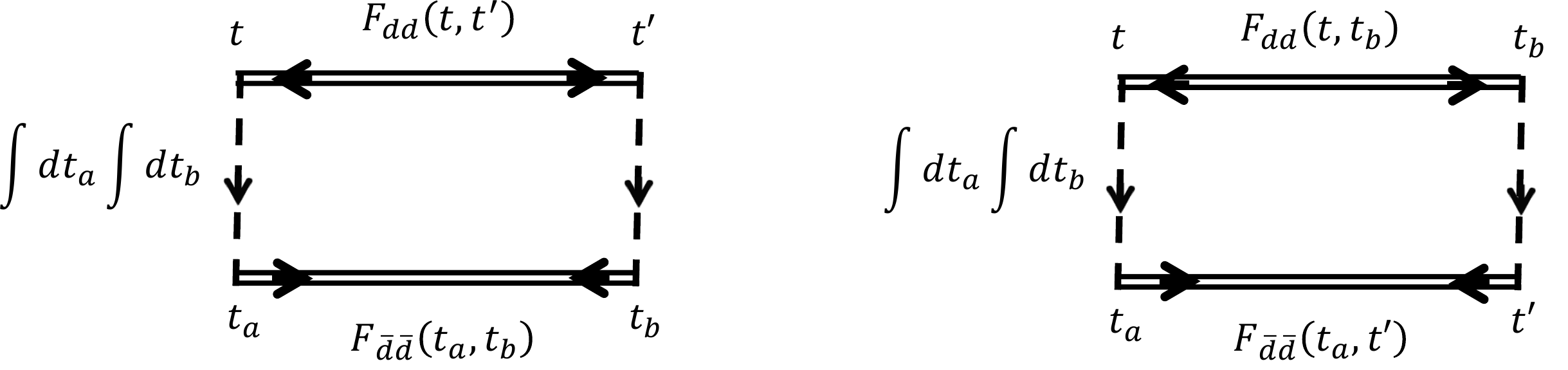}
\caption{ The anomalous contribution to the shot noise can be described by these two types of diagrams.
The left diagram (along with a similar diagram with arrow direction change) corresponds
to the first and the second lines in Eq. (\ref{eq:SN_anomalous1}), the right diagram
(along with a similar diagram with arrow direction change)
corresponds to the third and fourth lines in Eq. (\ref{eq:SN_anomalous1}).
}
\label{fig:SNDiaAno}
\end{figure}

We then consider the anomalous part $S_{I,A}(eV)$ in Eq. (\ref{eq:SNanomalous}). We first write down
equations of motion such that all Green function can be described by impurity Green function and
free electron Green functions:
\begin{eqnarray}
 F_{Lk,Lk'}(\omega) &=& -t_{k} t_{k'} g_{Lk}^0(\omega) F_{d,d}(\omega) \widetilde{g}_{Lk'}^0(\omega) ,\\
 F_{\bar{Lk},\bar{Lk}'}(\omega) &=& -t_{k}^{*} t_{k'}^{*} \widetilde{g}_{Lk}^0(\omega) F_{\bar{d},\bar{d}}(\omega) g_{Lk'}^0(\omega) ,\\
 F_{Lk,d}(\omega) &=& t_{k} g_{Lk}^0(\omega) F_{d,d}(\omega) , \\
 F_{\bar{Lk},\bar{d}}(\omega) &=& -t_{k}^{*} \widetilde{g}_{Lk}^0(\omega) F_{\bar{d},\bar{d}}(\omega) , \\
 F_{d,Lk}(\omega) &=& -t_{k} F_{d,d}(\omega) \widetilde{g}_{Lk}^0(\omega),  \\
 F_{\bar{d},\bar{Lk}}(\omega) &=& t_{k}^{*} F_{\bar{d},\bar{d}}(\omega) g_{Lk}^0(\omega).
\end{eqnarray}
We then insert those relations into Eq. (\ref{eq:SNanomalous}), and obtain
\begin{eqnarray}
  S_{I,A}(eV)&=& \frac{1}{2} \Big(\frac{e}{\hbar}\Big)^2 \int \frac{d\omega}{2\pi}
   \mathbf{Tr}\Big\{- \sum_k |t_{Lk}|^2 g_{Lk}^0(\omega) F_{d,d}(\omega) \sum_{k'} |t_{Lk'}|^2 \widetilde{g}_{Lk'}^0(\omega)
            \hat{\gamma}^q F_{\bar{d},\bar{d}}(\omega) \hat{\gamma}^q  \nonumber\\
   &&\quad\quad\quad\quad\quad\quad\quad\quad - F_{d,d}(\omega) \hat{\gamma}^q \sum_{k'} |t_{Lk'}|^2 \widetilde{g}_{Lk'}^0(\omega)
            F_{\bar{d},\bar{d}}(\omega) \sum_{k} |t_{Lk}|^2 g_{Lk}^0(\omega) \hat{\gamma}^q \nonumber\\
   &&\quad\quad\quad\quad\quad\quad\quad\quad + \sum_k |t_{Lk}|^2 g_{Lk}^0(\omega) F_{d,d}(\omega)\hat{\gamma}^q
                       \sum_{k'} |t_{Lk'}|^2 \widetilde{g}_{Lk'}^0(\omega) F_{\bar{d},\bar{d}}(\omega)\hat{\gamma}^q \nonumber\\
   &&\quad\quad\quad\quad\quad\quad\quad\quad +  F_{d,d}(\omega) \sum_k |t_{Lk}|^2 \widetilde{g}_{Lk}^0(\omega) \hat{\gamma}^q
                       F_{\bar{d},\bar{d}}(\omega) \sum_{k'} |t_{Lk'}|^2 g_{Lk'}^0(\omega) \hat{\gamma}^q \Big\} .
   \label{eq:SN_anomalous1}
\end{eqnarray}
In the diagrammatic representation, the anomalous contribution to the shot noise can be described by
the diagrams shown in Fig. \ref{fig:SNDiaAno}.
We insert the Green functions $F_{d,d}(\omega)$, $F_{\bar{d},\bar{d}}(\omega)$,
$g_{\alpha k}^0(\omega)$, and $\widetilde{g}_{\alpha k}^0(\omega)$ from Eq. (\ref{eq:QddGreen}), (\ref{eq:gPGreen}), and (\ref{eq:gHGreen})
into Eq. (\ref{eq:SN_anomalous1}).
In the $T=0$ limit, we simplify the the anomalous part of the shot noise and obtain
the result in the main text:
\begin{equation}
 S_{I,A}(eV)=\frac{2 e^2}{h} \int\limits_{-eV/2}^{eV/2} \mathbb{A}_{A}(\omega) d\omega,
\end{equation}
where
\begin{eqnarray}
  \mathbb{A}_{\rm A}(\omega) &=& \Gamma_L^2 \Big[ \big( F_{dd}^R(\omega) +F_{dd}^A(\omega) \big)^2
                            -8(\Gamma_L^2-\Gamma_R^2)
                             \frac{|F_{dd}^R(\omega)|^2}{\Sigma(\omega)} \big( F_{dd}^R(\omega) +F_{dd}^A(\omega) \big) \nonumber\\
                       &&\quad\quad +16(\Gamma_L-\Gamma_R)^2((\Gamma_L+\Gamma_R)^2+\epsilon_d^2)
                           \frac{|F_{dd}^R(\omega)|^4}{\Sigma(\omega)^2}\Big].
\end{eqnarray}

\subsection{The effect of the dot energy level $\epsilon_d$ for the spinless non-interacting model}

\begin{figure}[htp]
\centering
\includegraphics[width=7.0in,clip]{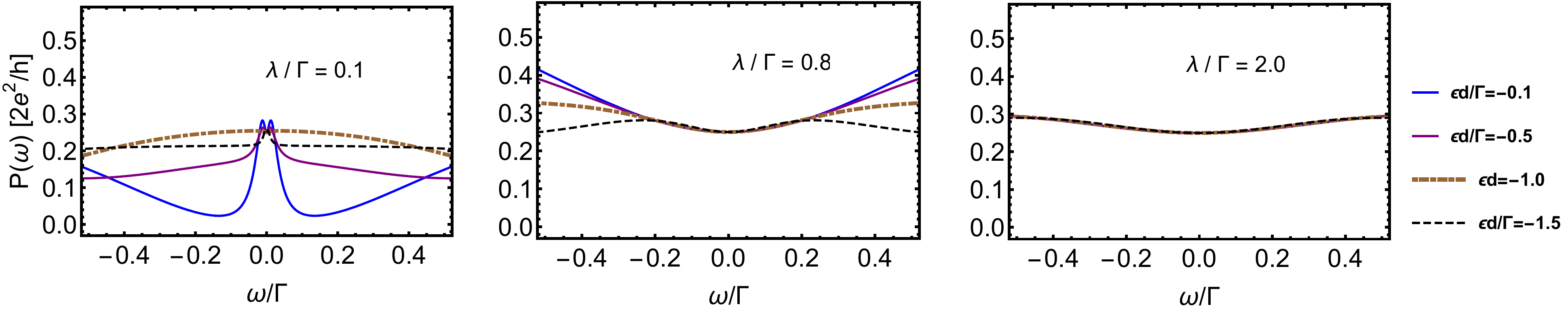}
\caption{ The power spectrum $P(\omega)$ (in units of $2e^2/h$) for (a) $\lambda/\Gamma=0.1$, (b) $\lambda/\Gamma=0.8$, and (c) $\lambda/\Gamma=2.0$.
We choose  $\Gamma_L=\Gamma_R$, $\delta=0.0$, and $\epsilon_d/\Gamma=-0.1, -0.5, -1.0, -1.5$.
}
\label{fig:Spectrum_varyEd}
\end{figure}

\begin{figure}[htp]
\centering
\includegraphics[width=5.0in,clip]{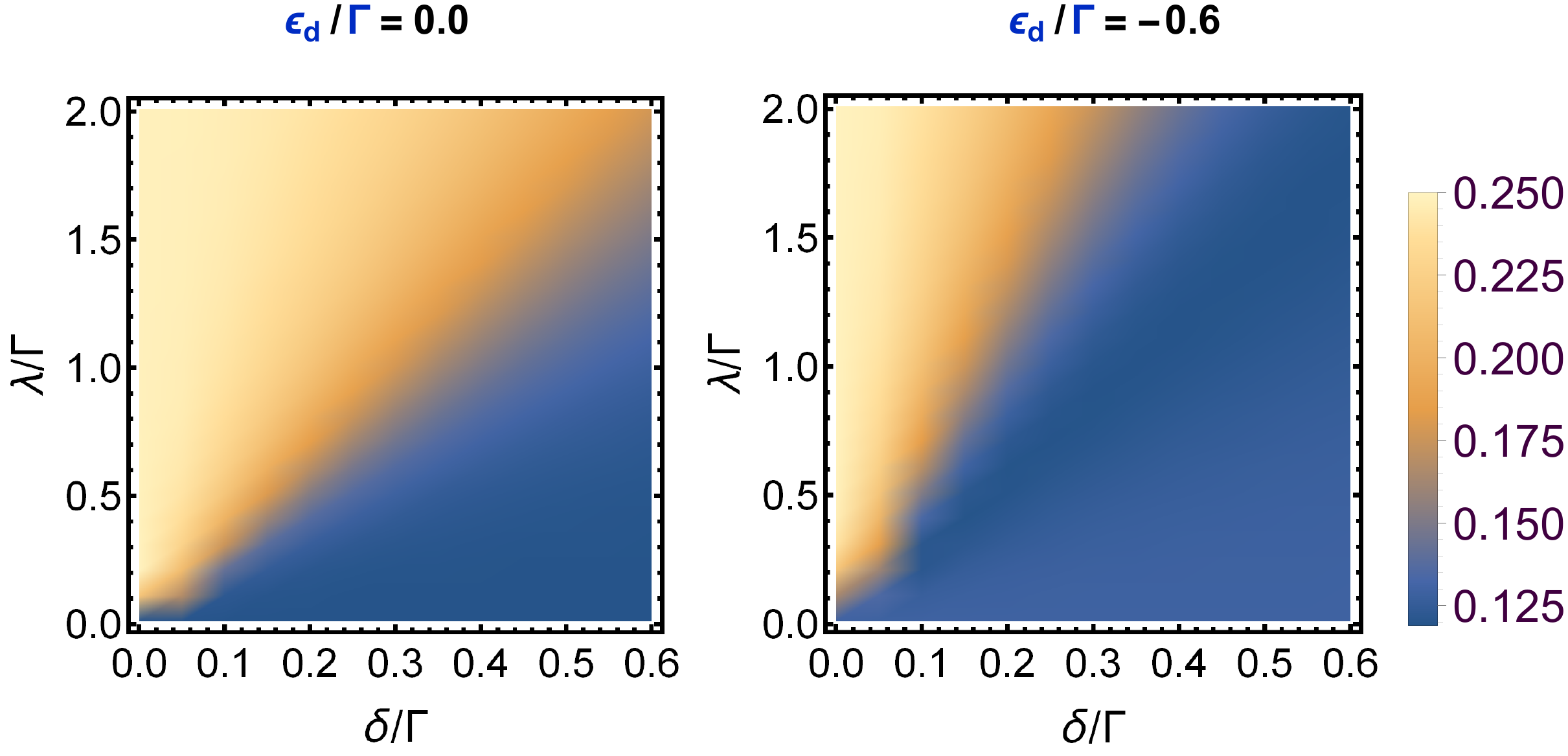}
\caption{ The finite bias shot noise $S_I(eV)/eV$ (in units of $2e^2/h$) as a function of $\lambda$ and $\delta$.
Left panel: $\epsilon_d=0.0$; right panel: $\epsilon_d/\Gamma=-0.6$.
We choose  $\Gamma_L=\Gamma_R$ and $eV/\Gamma=0.1$.
}
\label{fig:SNvsLamDel_twoEd}
\vspace{0.1in}
\end{figure}

Although the zero bias shot noise doesn't depend on the change of the QD chemical potential $\epsilon_d$,
i.e. $P(\omega)=\mathbb{A}_{\rm N}(\omega)+\mathbb{A}_{\rm A}(\omega)$ is independent of $\epsilon_d$ at $\omega=0$,
the change of $\epsilon_d$ can affect finite bias noise. The spectral function $P(\omega)$ for different $\epsilon_d$
and $\lambda$ are shown in Fig. \ref{fig:Spectrum_varyEd}. For small $\lambda$,
the shape and the width of the central peak show large changes under varying $\epsilon_d$.
For $\lambda\sim\Gamma$ and large $\lambda$, the width of the central regime
becomes flatter. We plot the shot noise for a small finite bias $eV/\Gamma=0.1$
as functions of $\lambda$ and $\delta$ in Fig. \ref{fig:SNvsLamDel_twoEd}, and compare the $\epsilon_d=0$
result with the  $\epsilon_d/\Gamma=-0.6$ result. The shot noise
$S_I(V)/eV$ shows crossover from non-universal value to $(e^2/2h)$ as $\lambda$ becomes large.
As the dot energy $|\epsilon_d|$ increases, the crossover line shifts to the position with larger $\lambda$.

\subsection{The Slave boson mean field approach}

We outline here the main steps of SBMF approach, more details can be found in Ref.~\cite{ChengPRX2014}.
Following standard procedure~\cite{ColemanPRB83,LeeRMP06}, one can introduce
the auxiliary boson $b$ and fermion $f_{\sigma}$ to replace the impurity operator by $d_{\sigma}\rightarrow f_{\sigma}b^{\dagger}$,
with the constraint $b^{\dagger}b+\sum_{\sigma}f_{\sigma}^{\dagger}f_{\sigma}=1$.
The Hamiltonian as shown in Eq. (1) of the main text becomes
\begin{eqnarray}
 H_{\rm SBMF} &=& H_{\rm Leads}+ \sum_{\sigma} \epsilon_d f_{\sigma}^{\dagger} f_{\sigma}
                        ++i\lambda \gamma_1 (f_{\uparrow}b^{\dagger}+f_{\downarrow}^{\dagger}b)
            +\sum_{\alpha=L,R}\sum_{k,\sigma} t_{\alpha} (c_{k\sigma,\alpha}^{\dagger} f_{\sigma}b^{\dagger}+h.c.)
             \nonumber\\
             &&+ i\delta\gamma_1 \gamma_2
              +\eta (b^{\dagger}b+\sum_{\sigma}f_{\sigma}^{\dagger}f_{\sigma}-1).
\end{eqnarray}
where the lead Hamiltonian $H_{\rm Leads}$ is unchanged, and the last term is Lagrangian multiplier
which enforces the constraint on the Hilbert space.

\begin{figure}[htp]
\centering
\vspace{0.2in}
\includegraphics[width=5.0in,clip]{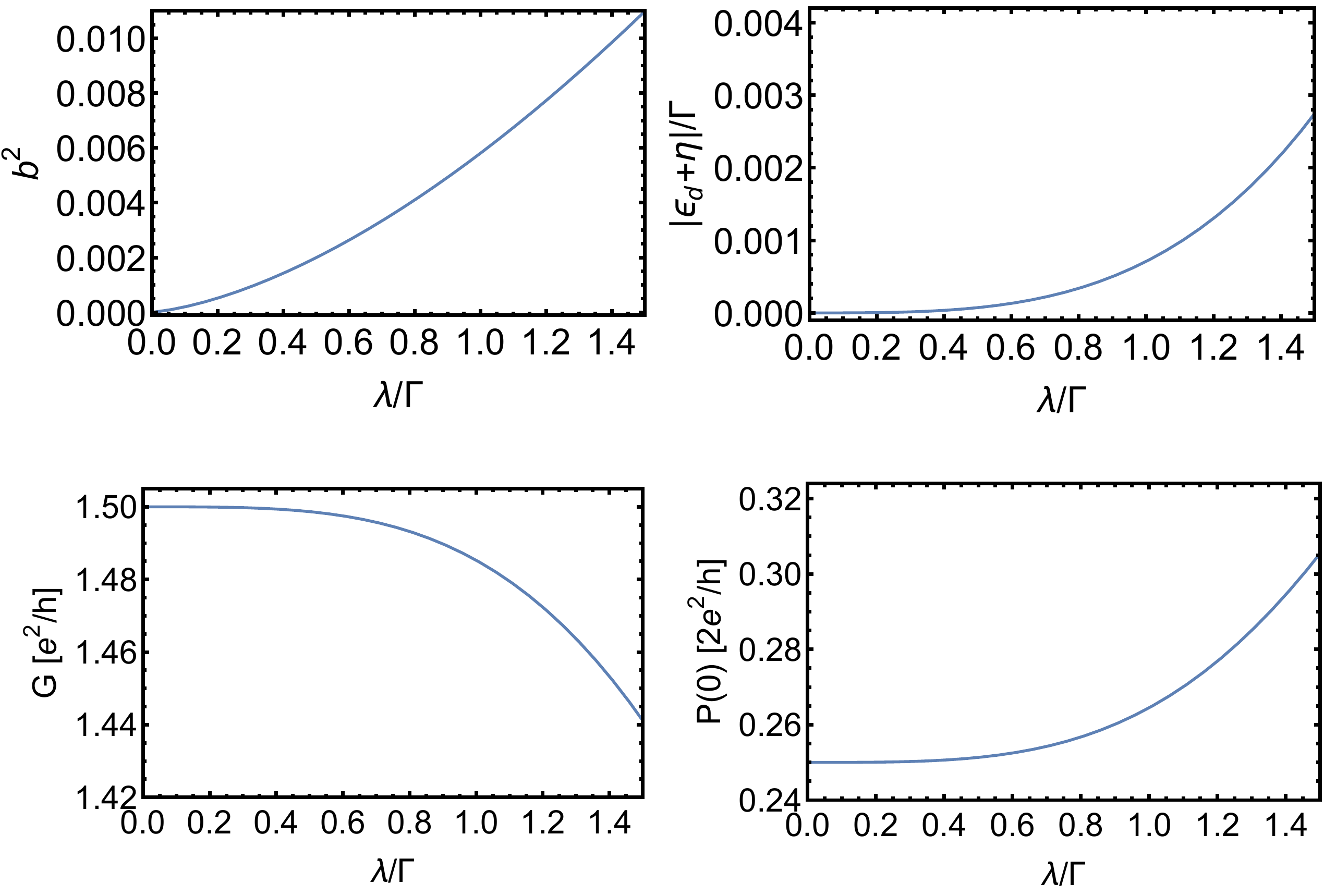}
\caption{Dependence of various parameters on the Majorana coupling $\lambda$. (a): The effective coupling in $\Gamma b^2$.
(b): The renormalized energy level for $|\epsilon_d+\eta|$.
(c) and (d): The SBMF result for the linear conductance (c) and for the shot noise
$P(0)$. The starting point of $\lambda$ is a very small non-zero number.
We choose $\Gamma_L=\Gamma_R$, $\epsilon_d/\Gamma=-10.0$, and band width $\Lambda/\Gamma=30.0$.
}
\label{fig:SBMFT_eV0_MFpara}
\vspace{0.2in}
\end{figure}

We apply mean field approximation and
replace the bosonic field and the Lagrangian multiplier by their expectation values. Because of
a $U(1)$ gauge invariance, we choose $\langle b \rangle=\langle b^{\dagger} \rangle=b$ to be
a real positive number. The mean field parameter $b$ and $\eta$ can be determined
self-consistently by minimizing the free energy, and yield the saddle point relations \cite{ChengPRX2014}:
\begin{eqnarray}
 && b^2 +\sum_{\sigma} \langle f_{\sigma}^{\dagger} f_{\sigma} \rangle = 1,\\
 && 2b\eta + t \sum_{\alpha=L,R} \sum_{k\sigma} (\langle f_{\sigma}^{\dagger}c_{k\sigma,\alpha} \rangle + c.c.)
  +i\lambda \Big\langle \gamma_1 (f_{\uparrow}^{\dagger}+f_{\uparrow}) \Big\rangle = 0.
\end{eqnarray}
Here, we assume the $eV\ll {\rm max} \,\{T_K,\lambda\}$ and thus neglect the dependence of the $eV$ in the SMBF calculations.
The effective coupling $\Gamma b^2$ and the renormalized energy level
$|\epsilon_d+\eta|$ as a function of $\lambda$ are shown in Fig. \ref{fig:SBMFT_eV0_MFpara} (a) and (b).
For $\delta=0$, we plot the SBMF results for the linear conductance and  the shot noise (in the limit $eV\rightarrow 0$)
in Fig. \ref{fig:SBMFT_eV0_MFpara} (c) and (d), which show the crossover from universal values to non-universal ones
as increasing the coupling $\lambda$.
Note that the $1/4$ values at even very tiny QD-MF coupling $\lambda$ is attributed to the $eV=0$ limit.
As shown in the main text, for $\lambda\ll b\Gamma$, the requirement to observe those half quantized values
is $eV\ll \lambda^2/\Gamma$.

In the discussion of the main text, we focus on the single-occupancy regime $|\epsilon_d|\gg\lambda,\Gamma$.
Beyond this limit, the mean field parameter $b$ becomes large, and the effective energy level
$\widetilde{\epsilon}_{d}=|\epsilon_d+\eta|$ shifts away from Fermi level.
In this case, although the energy level shift does not affect the universal values
(both linear conductance and shot noise)
for spin-up channel (due to MZM coupling), this level shift will affect the spin-down
channel if $|\epsilon_d+\eta|>\Gamma b^2$.
The linear conductance and shot noise can be summarized as follows. The linear conductance at $\Gamma_L=\Gamma_R$ reads
\begin{equation}
 G = \left\{
  \begin{array}{l l}
    \frac{e^2}{h} (\frac{1}{2}+1)=\frac{3e^2}{2h} & \; \text{for $|\epsilon_d+\eta|\ll\Gamma b^2$},\\
    & \\
    \frac{e^2}{h}\Big( \frac{1}{2}+\frac{(\Gamma b^2)^2}{(\epsilon_d+\eta)^2+(\Gamma b^2)^2}\Big) & \; \text{otherwise},
  \end{array} \right.
  \label{eq:G_compare}
\end{equation}
which is consistent with the numerical renormalization group calculation \cite{LeePRB13}.
The shot noise ($V\rightarrow 0$ limit and $\Gamma_L=\Gamma_R$) reads
\begin{equation}
 P(0) = \left\{
  \begin{array}{l l}
    \frac{2e^2}{h} ( \frac{1}{4} + 0) & \; \text{for $|\epsilon_d+\eta|\ll\Gamma b^2$},\\
    & \\
    \frac{2e^2}{h} \Big( \frac{1}{4} +\frac{(\Gamma b^2)^2 (\epsilon_d+\eta)^2}{((\epsilon_d+\eta)^2+(\Gamma b^2)^2)^2}\Big) & \; \text{otherwise}.
  \end{array} \right.
  \label{eq:SN_compare}
\end{equation}
Here, the first term in the bracket corresponds to the spin-up channel (with MZM),
and the second term corresponds to the spin-down channel (without MZM).

Note that Majorana zero modes also emerge in certain quantum impurity models ~\cite{Emery&Kivelson,Affleck93TCK,Affleck95TIK}.
i.e. two channel Kondo models \cite{Emery&Kivelson,Affleck93TCK} and two impurity Kondo models \cite{Affleck95TIK}.
For example, the half-quantized value of the conductance was also predicted in a double quantum dot model at
the two channel quantum critical point \cite{Zarand06TCK}. However, this requires fine-tuning in contrast to 
the Majorana physics discussed here, which is a robust phenomenon protected by the gap of a topological superconductor. \cite{GGTCK,Mebrahtu13}.

\subsection{Quantum dot with side-coupled non-Majorana bound states: analysis of false-positive signatures}

There are several false-positive explanations for the observed zero-bias peak in the 
Majorana experiments~\cite{Mourik2012, Das2012, Deng2012, Fink2012, Churchill2013}:  
1) impurity scattering in the lead close to the interface with the superconductor 
(i.e. enhancement of the Andreev conductance due to coherent backscattering to the interface)\cite{Bagrets12,Pikulin12,Neven13};
2) disorder in the topological superconducting wire \cite{liujie12, Neven13, LobosPRL13, Sau&DasSarma13, Hui14}. 
In the recently grown epitaxial interfaces between an s-wave superconductor and a semiconductor~\cite{MarcusHardGap}, 
the amount of interfacial disorder has been decreased significantly, and the background conductance (so-called "soft gap") 
problem has been solved. One can, however, imagine that some impurities close to the QD might 
lead to a few fermionic or Andreev bound states (ABS). 
In the following subsections, we analyze the shot noise in the QD coupled to non-Majorana bound states and 
show that the case with MZM has unique signatures which are qualitatively different from those with
non-Majorana bound states. Below, we present results for the shot noise through a QD in three different realistic cases: a) QD coupled to a spinless fermionic mode, b) QD is coupled to an Andreev Bound State and c) QD strongly hybridized with an s-wave superconductor, and discuss how one can identify MZM and distinguish between the Majorana and three aforementioned scenarios.

%As we shown in the main text, the shot noise through a QD at $\Gamma_L=\Gamma_R$ is given by a universal value independent of
%microscopic parameters due to the coupling to MZM.

As explained in the main text, we propose to measure both conductance and shot noise and compare different cases. The idea is very simple - 
one should first tune the conductance to its maximum value by adjusting, for example, $\Gamma_L/\Gamma_R$ or gate voltage,
then ground the SC and measure the shot noise around that point. There are several qualitative features that distinguish between Majorana and non-Majorana physics: \\
- shot noise is at maximum for MZM and is at minimum in the other cases (see below). Thus, by changing the left-right asymmetry one should observe single peak vs. double peak structure in the shot noise; \\
- shot noise for QD coupled to the MZM is independent of microscopic parameters (quantum dot energy level, QD-MZM coupling);
for QD with non-Majorana modes, the shot noise depends on microscopic parameters.
Therefore, by tuning, for example, the quantum dot energy level, one can distinguish Majorana zero mode with other non-Majorana modes.

\subsubsection{Side-coupled spinless fermionic bound state}
We now consider a case when a disorder-induced spinless fermionic level is coupled to a QD. The Hamiltonian for the spinless energy level in QD coupled to a fermionic level is given by
\begin{equation}
 H_{\rm Dot} = \epsilon_d d^{\dagger}d + \epsilon_f f^{\dagger}f + \lambda (d^{\dagger}f+f^{\dagger}d),
\end{equation}
Here $d$ and $f$ are fermionic annihilation operators in QD and disorder-induced level, respectively, with $\lambda$ being the hybridization. We assume that left and right  normal leads only couple to the QD. In the linear response regime, the shot noise is given by
\begin{equation}
 P(0) = \frac{2e^2}{h} \frac{4\Gamma_L \Gamma_R \epsilon_f^2 [(\lambda^2-\epsilon_d\epsilon_f)^2+\epsilon_f^2 (\Gamma_L-\Gamma_R)^2]}
                   {[(\lambda^2-\epsilon_d\epsilon_f)^2+\epsilon_f^2 (\Gamma_L+\Gamma_R)^2]^2}=\left. \frac{2e^2}{h} \frac{\Gamma^2 \epsilon_f^2 [(\lambda^2-\epsilon_d\epsilon_f)^2]}
                   {[(\lambda^2-\epsilon_d\epsilon_f)^2+\epsilon_f^2 \Gamma^2]^2} \right|_{\Gamma_L=\Gamma_R}.
\end{equation}
One can notice that $P(0)$ depends on the microscopic details whereas in the case of coupling to MZM $P(0)$ is independent
of the microscopic details and is given by the universal expression, see Eq. (16). Furthermore, when the modified QD level
is on resonance (i.e. $\lambda^2-\epsilon_d\epsilon_f=0$), the shot noise is vanishing contrary to the Majorana case where
it is finite constant. Thus, we conclude that one can distinguish between coupling to MZM vs. fermion level.

\subsubsection{Side-coupled Andreev bound state}
Let us now consider a spinless Andreev bound state (ABS) coupled to a QD. The Hamiltonian for the QD-ABS system can be written as
\begin{equation}
 H_{\rm Dot} = \epsilon_d d^{\dagger}d + \epsilon_f f^{\dagger}f + \lambda \Big(d^{\dagger} (u f^{\dagger} + v f) +h.c.  \Big).
\end{equation}
where $f$ and $f^\dag$ are quasiparticle annihilation and creation operators for the ABS. Here $u$ and $v$ are Bogoliubov amplitudes satisfying $|u|^2+|v^2|=1$.

The shot noise in the linear response regime reads
\begin{equation}
 P(0) = \frac{2e^2}{h} \frac{4\Gamma_L \Gamma_R \epsilon_f^2 [((u^2-v^2)\lambda^2+\epsilon_d\epsilon_f)^2+\epsilon_f^2 (\Gamma_L-\Gamma_R)^2]}
                   {[((u^2-v^2)\lambda^2+\epsilon_d\epsilon_f)^2+\epsilon_f^2 (\Gamma_L+\Gamma_R)^2]^2}=\left. \frac{2e^2}{h} \frac{\Gamma^2 \epsilon_f^2 [((u^2-v^2)\lambda^2+\epsilon_d\epsilon_f)^2]}
                   {[((u^2-v^2)\lambda^2+\epsilon_d\epsilon_f)^2+\epsilon_f^2 \Gamma^2]^2} \right|_{\Gamma_L=\Gamma_R}.
\end{equation}
One can notice once again that $P(0)$ depends on microscopic details and vanishes on resonance $(u^2-v^2)\lambda^2+\epsilon_d\epsilon_f=0$. In this sense, the shot noise in the presence of ABS is qualitatively similar to the spinless fermionic level.

\subsubsection{Superconducting order on QD due to proximity effect}
As a last example, we consider an effect of proximity induced superconductivity in the QD due to the large QD-SC coupling. The effective Hamiltonian for the QD is then given by
\begin{equation}
 H_{\rm Dot} = (\epsilon_d-\frac{V_z}{2}) d_{\uparrow}^{\dagger}d_{\uparrow}+(\epsilon_d+\frac{V_z}{2}) d_{\downarrow}^{\dagger}d_{\downarrow}
                +\Delta_P (d_{\uparrow}^{\dagger} d_{\downarrow}^{\dagger}+h.c.).
\end{equation}
where $\Delta_P$ is the induced superconducting pairing potential. Here we have neglected the charging energy as it is strongly renormalized by the coupling to the superconductor, see, e.g. Ref.\cite{lutchyn2007}. Assuming that the Zeeman splitting $V_z$ is large enough so that the spin degeneracy is lifted $V_z~\sim \Delta_P \gg \Gamma, eV, T$, the shot noise through the QD in the linear response regime at $\Gamma_L=\Gamma_R$ becomes
\begin{eqnarray}
% P_{\uparrow}(0) &=& \frac{2e^2}{h} \frac{4\Gamma_L \Gamma_R (\epsilon_d+V_z/2)^2 [(\lambda^2+(\epsilon_d-V_z/2)(\epsilon_d+V_z/2))^2+(\epsilon_d+V_z/2)^2 (\Gamma_L-\Gamma_R)^2]}
%                   {[(\lambda^2+(\epsilon_d-V_z/2)(\epsilon_d+V_z/2))^2+((\epsilon_d+V_z/2))^2 (\Gamma_L+\Gamma_R)^2]^2},\\
% P_{\downarrow}(0) &=& \frac{2e^2}{h} \frac{4\Gamma_L \Gamma_R (\epsilon_d-V_z/2)^2 [(\lambda^2+(\epsilon_d+V_z/2)(\epsilon_d-V_z/2))^2+(\epsilon_d-V_z/2)^2 (\Gamma_L-\Gamma_R)^2]}
%                   {[(\lambda^2+(\epsilon_d+V_z/2)(\epsilon_d-V_z/2))^2+((\epsilon_d-V_z/2))^2 (\Gamma_L+\Gamma_R)^2]^2}.
 P_{\uparrow}(0) &=& \frac{2e^2}{h} \frac{\Gamma^2 (\epsilon_d+V_z/2)^2 (\Delta_P^2+(\epsilon_d-V_z/2)(\epsilon_d+V_z/2))^2}
                   {[(\Delta_P^2+(\epsilon_d-V_z/2)(\epsilon_d+V_z/2))^2+((\epsilon_d+V_z/2))^2 \Gamma^2]^2},\\
 P_{\downarrow}(0) &=& \frac{2e^2}{h} \frac{\Gamma^2 (\epsilon_d-V_z/2)^2 (\Delta_P^2+(\epsilon_d+V_z/2)(\epsilon_d-V_z/2))^2}
                   {[(\Delta_P^2+(\epsilon_d+V_z/2)(\epsilon_d-V_z/2))^2+((\epsilon_d-V_z/2))^2 \Gamma^2]^2}.
\end{eqnarray}
where $P_{\sigma}(0)$ is the corresponding contribution to the shot noise for each spin channel. As one can see, the shot noise depends on microscopic parameters. In particular, at the gate voltage $\epsilon_d-V_z/2=0$, the expression for $P(0)$ reads $P(0) \sim \frac{2e^2}{h}\frac{\Gamma^2 V_z^2}{\Delta_P^4}$ and is small since $\Gamma \ll V_z\sim \Delta_P$. Thus, similar to both fermionic bound state and ABS, the shot noise in this case is given by a non-universal expression and is vanishing on resonance.

%\subsection{Distinguishing between MZM and other (non-topological) features}

\end{widetext}

\bibliography{ShotNoiseKondoMF,topological_wires11}

\end{document}